\documentstyle[12pt]{article}

\makeatletter
\@addtoreset{equation}{section}
\makeatother
\def\theequation{\arabic{section}.\arabic{equation}}
\newcommand{\As}{\hspace{.3ex}A\mbox{\hspace{-1.3ex}/}\hspace{.3ex}}
\newcommand{\ds}{\hspace{.3ex}\partial\mbox{\hspace{-1.3ex}/}\hspace{.3ex}}
\newcommand{\ks}{\hspace{.3ex}k\mbox{\hspace{-1.3ex}/}\hspace{.3ex}}

\newcommand{\Cte}{\mbox{\rm const.}}
\newcommand{\Tr}{\,\mbox{\rm Tr}\,}
\newcommand{\Oh}{\mbox{\rm O}}

\def\si{\sigma}
\def\ep{\epsilon}
\def\b{\begin{equation}}
\def\e{\end{equation}}
\def\lb{\label}

\def\cd{\cdot}
\def\ov{\over}

\def\a{\alpha}
\def\be{\beta}

\def\L{\Lambda}
\def\l{\left}
\def\r{\right}
\def\ba{\begin{eqnarray}}
\def\ea{\end{eqnarray}}
\def\no{\nonumber}
\def\p{\partial}
\def\la{\lambda}
\def\D{{\cal D}}
\def\intx{\int\!d^2\!x}
\def\intz{\int\!d^2\!z}
\def\intxi{\int\!d^2\xi}
\def\intxy{\int\!d^2\!x\!\int\!d^2\!y}
 \def\intly{\int_{a}\!d^2\!y}
\def\intk{\int\!{d^2\!k\ov(2\pi)^2}} \def\intp{\int\!{d^2\!p\ov(2\pi)^2}}
\def\intlp{\int^{\L}\!{d^2\!p\ov(2\pi)^2}}
\def\La{{\cal L}}

\def\vp{\varphi}
\def\ef{\mbox{\scriptsize eff.}}

\def\c{\mbox{\scriptsize c}}

\def\Hig{\mbox{\scriptsize Hig.}}
\def\ano{\mbox{\scriptsize anom.}}
\def\ex{\mbox{\rm e}}

\def\ra{\rangle}

\def\gf{\gamma_5}
\def\ra{\rightarrow}
\def\ch{\chi}
\def\Am{A^{\mu}} \def\An{A^{\nu}} \def\AM{A_{\mu}}
\def\pmu{p_{\mu}} \def\pn{p_{\nu}}
\def\m{\mu}\def\n{\nu}\def\mn{\m\n}
\def\gmn{g_{\mn}}
\def\ka{\kappa}
\def\el{\ell}

\def\msG{\mbox{\scriptsize msG}}

\def\F{{\cal F}}
\def\de{\delta}
\def\De{\Delta}
\def\kap{\ka_+}
\def\phz{\phi_0}
\def\pho{\phi_1}
\def\kaz{\ka_0}
\def\kao{\ka_1}
\def\dez{\de_0}
\def\deo{\de_1}

\def\PV{\mbox{\scriptsize PV}}

\def\rhoc{\rho_{\c}}
\def\sic{\si_{\c}}
\begin{document}
\ifx\TwoupWrites\UnDeFiNeD\else\target{\magstepminus1}{11.3in}{8.27in}
	\source{\magstep0}{7.5in}{11.69in}\fi
\newfont{\fourteencp}{cmcsc10 scaled\magstep2}
\newfont{\titlefont}{cmbx10 scaled\magstep2}
\newfont{\authorfont}{cmcsc10 scaled\magstep1}
\newfont{\fourteenmib}{cmmib10 scaled\magstep2}
	\skewchar\fourteenmib='177
\newfont{\elevenmib}{cmmib10 scaled\magstephalf}
	\skewchar\elevenmib='177
\newif\ifpUbblock  \pUbblocktrue
\newcommand\nopubblock{\pUbblockfalse}
\newcommand\topspace{\hrule height 0pt depth 0pt \vskip}
\newcommand\pUbblock{\begingroup \tabskip=\hsize minus \hsize
	\baselineskip=1.5\ht\strutbox \topspace-2\baselineskip
	\halign to\hsize{\strut ##\hfil\tabskip=0pt\crcr
	\the\Pubnum\crcr\the\date\crcr}\endgroup}
\newcommand\YITPmark{\hbox{\fourteenmib YITP\hskip0.2cm
        \elevenmib Uji\hskip0.15cm Research\hskip0.15cm Center\hfill}}
\renewcommand\titlepage{\ifx\TwoupWrites\UnDeFiNeD\null\vspace{-1.7cm}\fi
	\YITPmark\vskip0.6cm
	\ifpUbblock\pUbblock \else\hrule height 0pt \relax \fi}
\newtoks\date
\newtoks\Pubnum
\newtoks\pubnum
\Pubnum={YITP/U-\the\pubnum}
\date={\today}
\newcommand{\frontpageskip}{\vspace{12pt plus .5fil minus 2pt}}
\renewcommand{\title}[1]{\frontpageskip
	\begin{center}{\Large \bf
#1}\end{center}\par}
\renewcommand{\author}[1]{\frontpageskip\par\begin{center}
	{\authorfont #1}\end{center}
	\nobreak
	}
\newcommand{\andauthor}{\frontpageskip\centerline{and}\author}
\newcommand{\authors}{\frontpageskip\noindent}
\newcommand{\address}[1]{\par\begin{center}{\sl #1}\end{center}\par}
\newcommand{\andaddress}{\par\centerline{\sl and}\address}
\renewcommand{\thanks}[1]{\footnote{#1}}
\renewcommand{\abstract}{\par\frontpageskip\centerline{\fourteencp Abstract}
	\vspace{8pt plus 3pt minus 3pt}}
\newcommand\YITP{\address{Uji Research Center \\
	       Yukawa Institute for Theoretical Physics\\
               Kyoto University,~Uji 611,~Japan\\}}
\thispagestyle{empty}
\pubnum{93-15}
\date{July 1993}
\titlepage
\baselineskip 24pt
\title{ Long-distance properties of frozen
 U(1) Higgs \newline and axially U(1)-gauged
 four-Fermi models \newline in $1+1$ dimensions
}
\author{Hisashi Yamamoto
\thanks{E-mail address: hisashi@yisun1.yukawa.kyoto-u.ac.jp}
}
\YITP
\baselineskip 19pt
\abstract{
We study the long-distance relevance of vortices (instantons)
in an $N$-component axially U(1)-gauged four-Fermi theory in $1+1$
dimensions, in which a naive use of $1/N$ expansion predicts the
dynamical Higgs phenomenon.
Its general effective lagrangian is found to be a frozen U(1)
Higgs model with the gauge-field mass term proportional to
an anomaly parameter ($b$).
The dual-transformed versions of the effective theory are represented
by sine-Gordon systems and recursion-relation analyses are
performed.
The results suggest that in the gauge-invariant scheme ($b=0$)
vortices are always relevant at long distances, while in non-invariant
schemes ($b>0$) there exists a critical $N$ above which the
long-distance behavior is dominated by a free massless scalar
field.
}

\newpage
\section{Introduction}
\indent

In a previous paper~\cite{YI} we have studied the finite-temperature
phase structure of the quasi (2+1)-dimensional four-Fermi  theory with a
global chiral U(1) symmetry, in order to get insight into the
infrared dynamics of quasi two-dimensional superconductors.
Remarkably we found two kinds of ordered phases,
one is of a Kosterlitz-Thouless (KT) type and another is of a
Nambu-Goldstone (NG) type.
It is now a natural question to ask whether a Higgs mechanism
operates or not if the model is coupled to an axial gauge field,
i.e.~if the axial symmetry is gauged.
This question may be well parallel to the analogous issue in
superconductivity, namely, if a Meissner effect takes place or not
when an external magnetic field is applied to quasi two-dimensional
superconductors.
It will not be too much to say that the qualitative long-distance
properties of a (2+1)-dimensional model at finite temperature is
essentially similar to those of a (1+1)-dimensional model at zero
temperature, in which a local gauge symmetry is easier to treat.

On the other hand, four-Fermi interactions have recently attracted
revived attentions in particle physics as a source of dynamical
symmetry breaking in the Standard model~\cite{Nambu}.
In this context top quark constitutes a strong-coupling four-Fermi
interaction and condensates to replace a fundamental Higgs boson.
The main ingredient of these arguments is based on a Nambu-Jona-Lasinio
mechanism~\cite{NJL} with a mean-field-type approximation.
Although some detailed analyses of field-theoretical aspects have
recently been given for non-gauged models~\cite{HHJKS,ZJ,HKK},
those for gauged models including topological effects has not yet
been addressed.
In 3+1 dimensions it is not straightforward to treat analytically
the many-body effect of topological excitations and one may rely on
numerical simulations on a lattice.
In such situations it would be instructive to study related
low-dimensional models with an abelian gauge symmetry,
where topological excitations are more tractable.
Although results themselves may somewhat depend on a dimensionality
and an abelian nature of the gauge symmetry, we expect to have
qualitative and field-theoretical lessons which may prove useful for
studying more realistic, higher-dimensional models with (non-abelian)
gauge symmetries.

In view of these backgrounds, we wish to argue in the present article
the \newline non-perturbative long-distance properties of an
$N$-component chiral U(1) four-Fermi theory coupled to an
axial gauge field in $1+1$ dimensions~\cite{GN}, as a
simple model to discuss the dynamical Higgs phenomenon.
In standard notations the model is defined with the
lagrangian
\b
\La=\bar{\psi}\cd(i\ds-\si-i\gf\pi+\As\gf)\psi
-(N/2g^2)(\si^2+\pi^2)-(N/4e^2) F^2,    \lb{gff}
\e
where $g$ and $e$ are respectively a dimensionless four-Fermi coupling
and an axial-vector gauge coupling with a mass dimension, and a
dot $(\cd)$ means to take a sum over fermion species $i=1\sim N$
(Lorentz indices are suppressed).
$\La$ is invariant under the following axial gauge transformation:
\ba
     \psi(x) &\ra &  \ex^{i\gf\vp(x)}\psi(x), \no\\
     A(x)    &\ra &  A(x) + \p\vp,             \lb{axial}\\
(\si+i\pi)(x)&\ra &  \ex^{-2i\vp(x)}(\si+i\pi)(x). \no
\ea
In their classic paper~\cite{GN}, Gross and Neveu studied
the limit $N\ra\infty$ of (\ref{gff}) for fixed
$g^2$ and $e^2$.
For $A=0$, $\La$ is invariant under scale and global chiral
transformations.
In the limit $N\ra\infty$ the dynamical breaking of chiral
symmetry develops for all non-zero $g^2$, and $\si$ acquires a
vacuum expectation value (v.e.v.) $<\si>=\sic$, giving a mass to
the fermions (dimensional transmutation)~\cite{GN}.
In $1+1$ dimensions, however, this symmetry-broken solution,
which we call a {\it NG vacuum}, is unstable against
higher-order corrections of $1/N$ expansion.
Beyond the leading order, uncontrollable infrared divergences
appear in the gap equations (cancellation conditions of $\si$
tadpoles: $<\hat{\si}>=0$ where $\hat{\si}\equiv\si-\sic$)
due to the fluctuations of a massless NG boson $\pi$ (fig.1)
and render the leading-order solution meaningless for a general
finite $N$~\cite{Root}.
The absence of a stable NG vacuum is in accord with the
general no-go theorem~\cite{MWHC}.
The meaningful solution for the non-gauged model was later given
by Witten~\cite{Witten}, who argued that a stable solution
is provided by assuming a v.e.v. only for a modulus field $\rho$
($<\rho>=\rhoc$) of the order-parameter field
\b
            \si+i\pi=\rho\,\ex^{i\ch},   \lb{order}
\e
but not for a phase field $\chi$. At large $N$ the long-distance
behavior is described mainly by a free massless phase field, i.e.
with the effective lagrangian
\b
\La_{\ch}=(N/8\pi)(\p\chi)^2. \lb{phaselag}
\e
Other terms for $\ch$ (only) are all of higher-derivative and are not
important at long distances.
The chiral U(1) symmetry is not broken ($<\ex^{i\ch}>=0$) but
realized in a KT phase~\cite{B,KT}.
It is then expected that a propagation of the massless $\ch$ field (KT
boson) in higher-order diagrams of $1/N$ expansion does not
give any infrared singularity but only supplies finite
renormalizations to the value of $\rhoc$ and to the effective
lagrangian (\ref{phaselag}).
The absence of infrared divergence can be confirmed explicitly
in the second-order gap equation (fig.2).
The fluctuation of $\rho$ is massive and would only provide
finite renormalization effects.

Thus the vacuum specified by
\b\rhoc\neq0, \quad\quad <\ex^{i\ch}>=0,   \lb{KT-vacuum}\e
which we call a {\it KT vacuum}, is expected to be stable in $1/N$ expansion.
For large $N$ the long-distance behavior should qualitatively
be similar to that in a low-temperature phase of the XY model.
With use of (\ref{phaselag}) it is characterized by the
power-law decay of a phase-wave correlation function
\b
<\ex^{-i\ch(x)}\ex^{i\ch(0)}>\quad
\longrightarrow^{\raisebox{.8ex}
{\hskip -9mm\mbox{\scriptsize $|x|\ra\infty$}}}
 {\Cte\ov |x|^{\Cte/N}},  \lb{XYcorrelation}
\e
where $\Cte$ stand for some positive constant numbers.

In ref.~\cite{GN} Gross and Neveu have also studied the $N\ra \infty$
limit of (\ref{gff}) with the axial gauge coupling ($A\neq0$).
Based on the NG vacuum ($\sic\neq0$) they computed the two-point
effective action for $A$ and $\pi$.
With a suitable gauge fixing they showed that both of their
propagators acquire poles at $p^2=\a(\equiv e^2/\pi)$, indicating
the dynamical Higgs mechanism.
As a result all (axial) charges are expected to be screened.
We should note that, in contrast to the global model, this simple
mean-field argument in the gauged model is {\it stable under
higher-order corrections} even in $1+1$ dimensions and hence its
picture is not to be criticized from the viewpoint of
large fluctuations, whatever the dimensions are;
there exists no dangerous massless NG particle generating
an infrared singularity to give a qualitative modification.

Then we would like to ask, to what extent this picture of a dynamical
Higgs phenomenon is correct.
A possible source to change the picture would be effects from
{\it topological excitations} missed in a $1/N$-perturbation
theory.
As is well known in dilute-instanton-gas
calculations~\cite{CDG,MT,RU,Polyakov}, topological excitations in
various models usually work toward disordering the systems and may
well destroy a naive mean-field argument {\it if they are actually
relevant at long distances}.

Another point that should be considered carefully is a subtlety with
respect to an axial-gauge invariance, which is also neglected
in the old argument~\cite{GN}.
An anomaly could generally emerge if the fermions are
integrated with a gauge non-invariant regularization.
In $1+1$ dimensions, however, this reflection manifest itself
on the effective action just as a contact term $A^2$ of the axial gauge
field~\cite{Johnson,RS,Jackiw} and does not bring out a theoretical
difficulty; a massive abelian gauge (Proca) theory is
quantum-mechanically consistent.
Hence, following recent studies of anomalous gauge
theories~\cite{JR,MS}, it is natural to consider that the quantum
theory of such a type possesses a hidden one-parameter degree of
freedom which cannot be uniquely determined from the theoretical
consistency.
Once such a mass term is allowed it should play an
important role in the long-distance physics.

We thus believe it important to study whether the naive argument
in favor of a dynamical Higgs mechanism could persist to hold or not
even after incorporating above two points.

In $(1+1)$-dimensional models with an abelian symmetry, the relevant
topological excitation is a vortex configuration in a euclidean
space-time (instanton).
As is suggested from our previous work~\cite{YI}, in order to
incorporate vortex configurations in the present four-Fermi theory
we should start from the KT vacuum (\ref{KT-vacuum}) keeping
a global chiral U(1) symmetry.
This can conveniently be achieved with a radial parametrization
of the order-parameter field (\ref{order})~\cite{Witten,YI}.
In sect.2 we will show that a general large-$N$ long-distance
effective lagrangian on the KT vacuum consists of a U(1) Higgs model
with a radial fluctuation frozen and of the possible gauge-field mass
term, the coefficient ($\equiv b$) of which is arbitrary.
We will thus study the long-distance properties of the effective
lagrangian in two cases separately, i.e.~in a gauge-invariant scheme
($b=0$, sect.3) and in non-invariant (anomalous) schemes
($b>0$, sect.4).

The (normal) U(1) Higgs model itself is an interesting physical system,
for example as effective theories of thin-film superconductors or a
helium superfluid threatened by a magnetic field, which in fact
contain vortex solutions and some early analyses have been
reported~\cite{HNHF}.
In our knowledges, however, a proper quantum-mechanical treatment
of the gauge-vortex dynamics has not yet appeared in references.
With the help of a dual transformation, we will develop in sect.3
a renormalization-group (RG) analysis.
We first formulate the partition function of the frozen U(1) Higgs
model with an ensemble of single-vortex configurations.
The system is then equivalent with that of many
vortices interacting with a massive scalar field.
This scalar-vortex system is generalized to incorporate an additional
vortex chemical-potential controlled by a fugacity parameter $y$, and
the system with a large chemical potential ($y\ll1$) can be described
in terms of a massive sine-Gordon (s-G) field theory.
In this local field theory, we can investigate interaction effects
among vortices using a recursion-relation method.
Although the validity of the analysis is restricted to a
small-fugacity regime ($y\ll1$), we expect to have an important
suggestion about the phase structure of the original model ($y=1$),
in particular whether there could exist the phase where vortices
are irrelevant and the mean-field treatment is justified at long
distances.

Then in sect.4 we will study the general effective
lagrangian derived in a gauge non-invariant scheme, i.e.
with a non-zero gauge-field mass term.
With the similar procedures to the above we will show that the system
is equivalent to that of vortices interacting with both massive and
massless scalar fields.
The system with a large chemical potential ($y\ll1$)
is represented by a coupled system of massless and massive s-G fields
and the recursion-relation analysis will be performed as well.

Sect.5 is devoted to summary and concluding remarks.
Appendix A includes some technical details for deriving
momentum-shell recursion equations.
In Appendix B is presented an approximate argument to support
partly the conjecture for the original $y=1$ system in the
$b>0$ scheme.
\section{Effective lagrangian of the axially \newline
U(1)-gauged four-Fermi theory}
$\quad$
In this section, based on the KT vacuum (\ref{KT-vacuum})
we derive for the axially U(1)-gauged four-Fermi model
(\ref{gff}) the large-$N$ long-distance effective lagrangian,
corresponding to (\ref{phaselag}) for the non-gauged model.
The obtained lagrangian will provide a basis on which we can later
investigate the long-distance relevance of vortex configurations.

In terms of the radial parametrization (\ref{order}) of the
order-parameter field, the four-Fermi system we will consider,
is described by the partition function
\b Z=\int \rho\D\rho\D\ch\D A\D\bar{\psi}\D\psi\,
     \exp \l\{i\intx\La^{\prime}\r\},   \lb{partition} \e
where a local gauge symmetry is not yet fixed.
In (\ref{partition}), shifting a modulus field as
$\rho+\rhoc$, we have
\ba
\La^{\prime}&=&\bar{\psi}\cd(i\ds-\rhoc+\As\gf)\psi -(N/4e^2)F^2
               -N\rhoc\,\rho-N\rho^2/2              \no\\
            & &-\rhoc\bar{\psi}\cd(\ex^{i\ch\gf}-1)\,\psi
               -\rho\,\bar{\psi}\cd\ex^{i\ch\gf}\,\psi\,+\,\Cte.
   \lb{gff2}
\ea
To the leading order of $1/N$ expansion, a gap equation
(a cancellation condition of $\rho$ tadpoles $<\rho>=0$,
corresponding to the diagrams in fig.3) reads
\b 0={\rhoc\ov g^2}-i\intk \Tr {1\ov \ks-\rhoc}.  \lb{gapeq} \e
With an ultraviolet cut-off scale $\L$, the lowest-lying-energy
solution is given by
\b
\rhoc=\ex^{-\pi/g^2}\L=\ex^{-\pi/g^2_R(\mu)}\mu,
\lb{gap}
\e
where a renormalized coupling $g^2_R$ is defined by
\b
g^{-2}=g^{-2}_R(\mu)+(2\pi)^{-1}\ln(\L^2/\mu^2) \lb{renormalization}
\e
with $\mu$ a renormalization mass scale.

For all positive region of $g^2$ the gap $\rhoc$ is positive and
gives fermions a RG-invariant mass satisfying a homogeneous RG equation
\b
\l[\,\mu(\p/\p\mu)+\be(g^2_R)(\p/\p g^2_R)\,\r]\rhoc=0,
\e
with a $\beta$ function
\b
\be(g^2_R)=\mu(\p/\p\mu)g_R^2(\mu)=-(1/\pi)(g_R^2)^2.
\e
Note that as is in the non-gauged model~\cite{Witten},
even if fermions acquire a mass through a non-zero $\rhoc$
the global chiral symmetry of full $\La^{\prime}$ is kept unbroken.
In other words, the present KT scheme of the four-Fermi dynamics drives
a dimensional transmutation but not a spontaneous breaking of the
global chiral symmetry.

Integrating over fermions we obtain the effective action to the
leading order in $1/N$.
The long-distance part is contained in the two-point
functions for $A$ and $\ch$ including their mixing (fig.4).
They read in momentum space as
\b
S_2=N\intp \,[\,\La_{\rm inv.}(p) +\La_{\rm anom.}(p)\,],  \lb{twopoint}
\e
with
\ba
\La_{\rm inv.}(p)
    &=&(2e^2)^{-1}\Am(-p)\,(\pmu\pn-\gmn p^2)\,(1+\a V)\,\An(p)  \no\\
    & & +(2\pi)^{-1}\rhoc^2U\,[\,\Am(-p)-(ip^{\mu}/2)\chi(-p)\,]\,
           [p \ra -p ],   \lb{inv}   \\
\La_{\rm anom.}(p)&=&(2\pi)^{-1}\,b\,\Am(-p) A_{\mu}(p), \lb{anomaly}
\ea
where $\Am(p)$ and $\ch(p)$ represent Fourier components
of $\Am(x)$ and $\ch(x)$ respectively.
Here $U$ and $V$ are functions of $p^2$ and $\rhoc^2$, defined by
\ba
U(p^2,\rhoc^2)&=&{2\ov\sqrt{p^2(p^2-4\rhoc^2)}}
\ln\l({\sqrt{-p^2+4\rhoc^2}+\sqrt{-p^2} \ov
\sqrt{-p^2+4\rhoc^2}-\sqrt{-p^2}}\r),  \no\\
V(p^2,\rhoc^2)&=&(\rhoc^2U-1)/p^2.     \lb{UV}
\ea
They have the following well-defined derivative expansions
in a configuration space:
\ba
\rhoc^2U(-\p^2,\rhoc^2)&=&1-\p^2/6\rhoc^2+\Oh((\p^2/\rhoc^2)^2),
\no\\
V(-\p^2,\rhoc^2)&=&(6\rhoc^2)^{-1}\,[\,1+\Oh(\p^2/\rhoc^2)\,],
\lb{derexp}
\ea
accompanied by inverse powers of $(\mbox{\rm mass})^2$ scales
$\rhoc^2$.
The effective action in (\ref{twopoint}) consists of two parts:
one is a locally (axial) gauge-invariant part ($\La_{\rm inv.}$) and
another is a non-invariant contact mass term ($\La_{\rm anom.}$)
originated from the ambiguity in the evaluation of a local
contribution in fermion loops, i.e.\ the axial-vector
anomaly~\cite{Johnson,RS,Jackiw}.
In (\ref{anomaly}) a parameter $b$, called an (hidden) anomaly
parameter in the literatures of anomalous gauge theories~\cite{JR,MS},
takes an arbitrary value, depending on a regularization procedure
for fermion loops.

Since an ultraviolet regularization is necessary only for
the $A_{\m}-A_{\n}$ vacuum polarization diagram (fig.4(a)),
it is sufficient, for example, to add to the
original lagrangian $\La^{\prime}$ the following Pauli-Villars
regulator part $\La_{\PV}$:
\b
\La_{\PV}=\bar{\psi}_1^{\PV}\cd(\,i\ds+b_1\As+\L\,)\,\psi_1^{\PV}
          +\bar{\psi}_2^{\PV}\cd(\,i\ds+b_2\As\gf+\L\,)\,\psi_2^{\PV},
\lb{PV}
\e
with $b_1^2+b_2^2=1$, the condition required for the cancellation of
apparent logarithmic divergences.
Of course, other regularization prescriptions such as a point-splitting
method would also be available.
In the present Pauli-Villars method, starting from
$\La^{\prime}+\La_{\PV}$, we in fact obtain (\ref{twopoint}) with $b$
in (\ref{anomaly}) given explicitly by
\b b=-1+b_1^2-b_2^2=2\,(b_1^2-1),               \lb{b} \e
which is an arbitrary constant.
\footnote
{
Although the hermiticity of $\La_{\PV}$ is lost except for
$0\leq b_1^2 \leq 1$, it would not lead to a serious matter because
$\psi_i^{\PV}$ are introduced only for a technical purpose,
i.e.\ only as regulator fields and disappear from the physical
spectrum $(\L\ra\infty$) after fermion-loop computations.
The remnant of the regulators only emerge as $\La_{\rm anom.}$.
}

$\La_{\rm inv.}$ in (\ref{inv}) is (axially) gauge invariant and
one should get only this part provided he imposes the gauge
invariance on the gauge-field vacuum polarization by employing,
for example, the above Pauli-Villars method with the special choice
$b_1=1\,(b_2=0)$.
Alternatively, for the practical purpose this gauge-invariant result
is most easily achieved by use of the dimensional regularization,
defined only in momentum integrals with all algebras of
$\gamma^{\mu}$ matrices unchanged from those of two-dimensional ones
(a dimensional reduction)~\cite{SCJN}, or that with an assumption
that $\gf$ anticommutes with all $\gamma^{\mu}$ matrices in continuous
$d$ dimensions~\cite{CFH}.
In this gauge-invariant system (\ref{inv}), if $\chi$ is single
valued, we can absorb it into a gauge-invariant vector field
\b
B^{\mu}(p)=\Am(p)+(ip^{\mu}/2)\ch(p),   \lb{vector}
\e
and a $\chi$ integration is decoupled from the theory.
Then, eq.(\ref{inv}) is reduced to a free part of the neutral
massive vector theory with the propagator $D_{\mu\nu}$
\b
iD_{\mu\nu}(p)= e^2(\gmn- (\pmu\pn/p^2))(p^2-\a)^{-1}
              -(\pi/\rhoc^2U)(\pmu\pn/p^2),
\lb{vecpro}\e
which possesses a pole at $p^2=\a$.
This mass $\a^{1/2}$ is independent of the four-Fermi coupling $g^2$
and is identical with that of the Schwinger model~\cite{Schwinger}.
Although our effective action (\ref{twopoint}) is gauge invariant
the feature of the spectrum is the same as that predicted in
the broken-symmetry ($\sic\neq0$) argument~\cite{GN}.
This may thus be called a gauge-invariant version of the dynamical
Higgs mechanism in a (1+1)-dimensional chiral four-Fermi
theory.

As is mentioned above, although $b$ can be fixed to be zero from the
gauge invariance, we are in principle free to choose it to suit our
requirements on the theory.
Since the massive abelian gauge theory is a quantum-mechanically
consistent system, any choice of $b(\geq0)$ would give a consistent
$1/N$ perturbation.
Keeping $b\geq 0$ arbitrary, we therefore consider our effective action
(\ref{twopoint}) as a {\it one-parameter family} of theories,
and will henceforth study the long-distance properties of this family.
This idea of the hidden-parameter generalization has been popular
in recent studies of anomalous gauge theories~\cite{JR,MS}.


To the leading order the two-point function of $\rho$ can also be
calculated with use of the gap equation (\ref{gapeq}).
The $\rho$ fluctuation is found to be short-ranged with a mass $2\rhoc$.
In higher orders the integration of this short-ranged fluctuation
would give finite corrections of $\Oh(N^0)$ to $\La_{\rm inv.}$.

Even for the gauge-invariant system $\La_{\rm inv.}$, physics seems
 more non-trivial than the above Higgs picture if $\ch$ is
{\it not} single valued.
In this case the field strength $F$ of $B$ will induce singularities
and the mean-field treatment of the dynamical Higgs mechanism
based on the replacement (\ref{vector}) appears to be too naive.
This possibility can generally happen since $\ch$ is an angle variable
of the axial rotation ((\ref{axial}), (\ref{order}))
and need not be single valued.
Such multi-valued configurations are physically realized as
topological excitations of the theory, that is, vortices in the
present abelian theory in $1+1$ dimensions.
To argue this aspect especially at long distances,
it is useful to extract the long-distance effective theory
contained in the gauge-invariant part $\La_{\rm inv.}$.
This effective theory should be described in terms of
a gauge-invariant local operator $F$ and a covariant one
$D\Psi$ ($D\equiv\p+iA$, $\Psi\equiv\ex^{i\chi}$),
with the following lagrangian:
\b
\La_{\Hig}=-(\be/4)F^2+(\ka/2)|D\Psi|^2,  \lb{higgs}
\e
where the gauge field has been rescaled as $\Am\ra\Am/2$ and
\ba
\be&\equiv&(N/4e_{\ef}^2)=(N/4e^2)\,[\,1+(\a/6\rhoc^2)+\Oh(1/N)\,],\no\\
\ka&\equiv&(N/4\pi)\,[\,1+\Oh(1/N)\,].           \lb{beka}
\ea
$\La_{\Hig}$ is nothing but a U(1) Higgs model with a radial
fluctuation frozen, or a continuum version of the XY model coupled to
a U(1) gauge field.
It possesses the same local gauge symmetry as the non-local action
$\La_{\rm inv.}$ and contains its lowest-derivative part for the
gauge and Higgs dynamics.
The higher-derivative terms which have been neglected include
$(\p F)^2$, $\p^2|D\Psi|^2$, etc.
Rescaling $\Am \ra e\Am$ we see both operators $F^2$ and $|D\Psi|^2$
possess canonical dimensions 2 and higher-derivative ones  4 and higher.
Hence at long distances the latter is irrelevant in a
RG point of view unless large anomalous dimensions are generated
dynamically, the possibility of which cannot be excluded but shall not
be considered here.
Although we have so far not considered four and higher-point functions
of $(A,\ch)$, it is straightforward to compute them.
{}From the gauge invariance the results should anyhow consist of only
higher powers of $F$ and $D\Psi$ like $(F^2)^2$ or $(|D\Psi|^2)^2$
than $\La_{\Hig}$ and also of their higher derivatives.
They are all irrelevant as seen from the naive dimensional counting.
{}From the universality argument, $\La_{\Hig}$ should thus
possess the same long-distance behavior as of $\La_{\rm inv.}$.

Our derivation of the effective lagrangian is based on $1/N$
expansion and hence, from its validity, parameters $\be$ and
$\ka$ defined by (\ref{beka}), should formally satisfy
\b
\be e^2 \gg 1, \quad\quad \ka \gg 1.    \lb{parameterregion}
\e
{}From general interests in $\La_{\Hig}$ itself, however, we shall
henceforth consider general regions of $\be e^2>0$ and $\ka>0$.
We expect that the qualitative properties at long distances
are same between the full effective theory $\La_{\rm inv.}$ and its
lowest-derivative part $\La_{\rm Hig.}$, for a general $N$ satisfying
$1\ll N \leq \infty$.

To summarize this section we have constructed the long-distance effective
lagrangian for the axially U(1)-gauged four-Fermi theory in $1+1$
dimensions, which reads
\b
\La_{\ef}=\La_{\Hig}+\La_{\rm anom.},
\quad \La_{\rm anom.}=(\,b\,N/8\pi) A^2, \quad\;b\geq0.  \lb{eff}
\e
\section{Long-distance properties of the frozen U(1)  \newline
         Higgs model and a massive s-G system}
$\quad$
In this section we consider the effective theory in the gauge-invariant
scheme ($b=0$), i.e.\ the frozen U(1) Higgs model defined with the
lagrangian $\La{\Hig}$ in (\ref{higgs}).
As does so in $\La_{\rm inv.}$, without singular configurations for a
phase order-parameter $\ch$ the Higgs mechanism operates
in this model;
the theory is reduced to that of a neutral massive vector boson
with a mass $M\equiv(\ka/\be)^{1/2}$ and any external (axial) U(1)
charge is predicted to be screened.
However, this naive picture could be changed once we allow
singular configurations for $\ch$.
In $1+1$ dimensions they are vortices in a euclidean space-time~\cite{NO}.
They are singular at their centers and cannot be absorbed into the
vector field by a non-singular gauge transformation (\ref{vector}).
%

The many-body dynamics of vortices in the (1+1)-dimensional
U(1) Higgs model (\ref{higgs}) has been argued in several early
references~\cite{CDG,MT,RU,ES,JKS}.
Two different approaches were mainly taken to attack the problem.
In one approach they relied upon a so called dilute-instanton-gas
approximation based on $\theta$ vacuums~\cite{CDG,MT,RU}.
The main results are the followings;
the energy density of the $\theta$ vacuums is proportional to
$-\cos\theta$ and the lowest-energy solution exists at $\theta=0$
where the global gauge symmetry is restored
(the v.e.v. of $\Psi$ is
zero due to the random varying of phase $\chi$):
the gauge-field propagator possesses a massless pole in addition
to the massive pole at $p^2=M^2$, and the external charge $Q$
is screened only if $Q$ is an integral multiple of $e$ and
is confined otherwise.
Thus, the long-distance picture in this semi-classical approximation
differs from the perturbative Higgs picture.

For these consequences to be justified, however, it still remains as an
important but unsolved problem {\it to verify dynamically
that, vortex excitations are indeed relevant and their distribution
is sufficiently dilute, at long distances}.
To solve this problem we must probe the long-distance
behavior of the system {\it quantum-mechanically}, in other words,
based on a {\it RG analysis} which has not yet properly been performed.

%
Another typical approach~\cite{ES,JKS} is to consider the theory on a
lattice and transform it to a system of a spin-wave (scalar field)
plus vortices with the help of a dual transformation~\cite{JKKN,BMK}.
In this approach also, although several possible phases were proposed
in a wide range of parameters, the determination of the true long-distance
behavior requires again a proper RG analysis which is not
straightforward on a real space-time lattice and, to our knowledge,
has never been undertaken.

In the rest of this section we will study the vortex dynamics of the
frozen U(1) Higgs model (\ref{higgs})
with the method being complementary to early investigations.
We keep the system (\ref{higgs}) in the continuum with a momentum
cut-off, below which our long-distance effective lagrangian is valid,
and take into account explicitly an ensemble of single-vortex
configurations in the partition function.
Then we shall proceed along the similar line of the lattice
approach~\cite{ES,JKS}.
Namely, we first transform the model to a scalar-vortex system by the
continuum version of a dual transformation~\cite{Sugamoto},
and next generalize it by adding a chemical potential of vortices,
controlled by a fugacity parameter $y$.
The small-$y$ regime $(0<y\ll1)$ of this generalized system can be
described by a local field theory, a massive s-G system (subsect.3.1).
In this system we will be able to study the long-distance relevance of
vortices in a systematic manner, i.e.\ by incorporating interaction
effect among vortices through a momentum-shell recursion-relation
method (subsect.3.2).
\subsection{Dual transformation in the cut-off continuum and a
 s-G system}
$\quad$
We start from the following euclidean partition function of the frozen
U(1) Higgs model (\ref{higgs}) with an ensemble of single-vortex
configurations:
\ba
Z&=& \prod_x \!\int\!\D A(x)\prod_{x_0}\sum_{n(x_0)}\prod_x\!\int\!
\D\ch(x)\delta \l[\,\ch(x)-n(x_0)\,\theta(x-x_0)\,\r]   \no\\
 & & \exp\l\{-\intx \l[\,{\be\ov4}F^2+{\ka\ov2}\,(A-\p\ch)^2 \,\r]\r\},
\lb{higgspf}
\ea
where $x_0$ are the positions of vortex centers and $\theta(x-x_0)$
denotes the azimuthal angle of $x$ around $x_0$.
In (\ref{higgspf}) the case $n=0$ corresponds to a usual unitary
gauge $(\ch=0)$. Here we sum up over all vortex charges $n(x_0)\in Z$
defined at all possible centers $x_0$.
The vortex centers are assumed to be separated from each other with
the minimal length $a\equiv\L^{-1}$ where $\L$ is the cut-off
scale we are considering; in our effective theory of the four-Fermi
model it is naturally given by the scale of order $\rhoc$.
Performing the continuum version~\cite{Sugamoto} of a dual
transformation~\cite{JKKN,BMK}
\b
\exp\l\{-{\be\ov4} \intx F^2 \r\}
=\prod_x\!\int\!\D\phi\exp\l\{-\intx\l[\,{1\ov2\be}\,\phi^2
  -i\phi\,(\ep\cd\p A)\,\r]\r\}  \lb{dual}
\e
with $\ep\cd\p A\equiv\ep_{\mu\nu}\p_{\mu}A_{\nu}\,(\ep_{12}=-\ep_{21}=1,
\ep_{11}=\ep_{22}=0$), and integrating over the gauge field $A$, we
obtain
\ba
Z&=&\prod_x\!\int\!\D\phi(x)\prod_{x_0} \sum_{n(x_0)}\!\prod_x\!
    \int\!\D\ch(x)\delta\l[\,\ch(x)-n(x_0)\,\theta(x-x_0)\,\r]   \no\\
 & & \exp\l\{-\intx \l[\,{1\ov2\ka}\,(\p\phi)^2+{1\ov2\be}\,\phi^2
             -2\pi i p \,\phi \, \r]\,\r\},
\lb{dualpf}
\ea
where
\b
p(x)\equiv(2\pi)^{-1}\ep_{\mn}\p_{\mu}\p_{\nu}\ch         \lb{p}
\e
is a vortex source.
The value of $p(x)$ is zero for a non-singular $\ch(x)$, but for our
singular $\ch(x)=n(x_0)\theta(x-x_0)$ imposed by the delta functional,
it takes $p(x)=n(x_0)\,\delta^2(x-x_0)$.
Then, integrating over the phase field $\chi(x)$ yields
\ba
Z&=&\prod_x \int\! \D\phi(x)\prod_{x_0}\sum_{n(x_0)\in Z} \no\\
 & & \exp\l\{-{1\ov2\ka}\intx \l[\,(\p\phi)^2+M^2\,\phi^2\,\r]
     +2\pi i \sum_{x_0}\, n(x_0)\,\phi(x_0) \r\},  \lb{swvor}
\ea
which describes the system of many vortices interacting with
a massive scalar field $\phi$.
The integration over $\phi(x)$ provides
\b
Z= Z_{\rm sca.}^{\rm free}\, \prod_{x_0}\sum_{n(x_0)\in Z} \exp\l\{
-2\pi^2\ka
\sum_{x_0,y_0} \,n(x_0)D(x_0-y_0;M)\,n(y_0) \,\r\},
\lb{SWvor}
\e
where $Z_{\rm sca.}^{\rm free}$ represents the partition function of the
free massive scalar field.
In the presence of gauge interaction $(\be<\infty)$ the interaction
between vortices is short-ranged
\b D(x-y;M)=\intlp\, {e^{ip\cdot(x-y)}\ov p^2+ M^2},
 \lb{SWpropagator}\e
in contrast to a long-range logarithmic one $D(x-y;0)$ for a pure
(non-gauged) XY model.
On the coincidence point $x=y$, one finds
\b D(0;M)=(4\pi)^{-1}\ln [\,(1+Z)/Z\,]
\e
with $Z\equiv M^2\L^{-2}$, which shows that a single vortex has a
finite energy so long as $M^2>0 (\be<\infty)$.
Thus, if we neglect the interaction of vortices, the vortex part of
the partition function is written by
\b
Z^{\rm free}_{\rm vor.}=\prod_{x_0}\sum_{n(x_0)}
\l({Z\ov 1+ Z}\r)^{{\pi\ka\ov2}\,n^2(x_0)}, \lb{freevor}
\e
from which the probability $P(x_0)$ to find a vortex
on a given position $x_0$ reads as
\b
P=\l({Z\ov 1+Z}\r)^{1+{X\ov2}}   \lb{probability}
\e
with $X\equiv\pi\ka-2$, and does the chemical potential as
\b
{\pi\ka\ov2}\ln{\l(1+Z\ov Z\r)}.    \lb{chemical}
\e

Eq.(\ref{SWpropagator}) shows  that
the interaction length $r$ among vortices
is roughly  $r \sim M^{-1} = Z^{-1/2}a$ and so the area in which a
vortex can affect is estimated by $(r/a)^2\sim Z^{-1}$ in the unit of
$a$.
Hence, from (\ref{probability}) the number of other vortices lying
closely enough to interact with a given vortex reads approximately
as
\b
n^{*}\equiv Z^{X\ov2}(1+Z)^{-(1+{X\ov2})}.  \lb{number}  \e
We expect this expression to be qualitatively valid also for a
smooth momentum cut-off.
Although this expression of $n^{*}$ is derived from a classical
picture we can also define an effective one
\b
n^{*}(\el)=Z(\el)^{X(\el)\ov2}(1+Z(\el))^{-(1+{X(\el)\ov2})},
\lb{effectiven}\e
for which quantum interaction effects are to be taken into account
through a RG analysis.
Here the dependences of $X(\el)$ and $Z(\el)$ on a scaling
parameter $\el$ are to be determined by RG.
The quantity $n^{*}(\el)$ thus changes dynamically as a function of
$\el$ and measures how vortices are effectively
dilute (or dense) at long distances ($\el \ra \infty$).
$P(\el)$ is defined similarly as well.

Now, we generalize the system (\ref{swvor}) by introducing an additional
chemical potential for a single vortex, the prescription having been
developed for the study of the XY model~\cite{JKKN}.
This leads us to
\ba
Z(y)&=&\prod_x \int\! \D\phi(x)\,\sum_{n(x_0)\in Z}
 \exp\l\{-\,{1\ov2\ka}\intx \l[\,(\p\phi)^2+M^2\,\phi^2\,\r] \r.\no\\
 & & \l.+(\ln y)\,\sum_{x_0}n(x_0)^2 + 2\pi i\,\sum_{x_0}n(x_0)\,\phi(x_0)
    \r\},   \lb{swvory}
\ea
where a chemical potential term $(\ln y)\, n^2(x_0)$ has been
incorporated by hand to control fluctuations in $n(x_0)$.
Although the original model corresponds to $y=1$,
we will henceforth consider this generalized system
$Z(y)$, regarding $y$ as a free parameter (a fugacity)
which controls the activation of single vortices.
In the generalized system the probability
$P$ of a single-vortex excitation  and the quantity $n^{*}$
defined respectively in (\ref{probability}) and (\ref{number})
are multiplied by $y$.
Our interests then lie in how this parameter $y$
behaves at long distances according to a RG and especially in
whether there exists the parameter region where $y(\el)$ vanishes in
the long-distance limit $(\el\ra\infty)$. If such a region exists,
there can we neglect the long-distance effect of vortices and the
mean-field picture ($n=0$) would be justified.
It is technically difficult to directly examine the RG behavior
in the general region of $y$.
However, it is interesting and feasible to consider the small-$y$
regime and to study how the system $Z(y)$ dynamically responds to
a compulsory reduction $(y\ll1)$ of vortex activation~\cite{JKKN}.

In the region $y\ll1$, the sum over vortex charges can be evaluated as
\ba
&&\ln\! \sum_{n(x_0)}\exp\l\{(\ln y)n^2(x_0)+2\pi i
   n(x_0)\,\phi(x_0)\r\}\no\\
&=&\ln \l\{1+2\sum_{n(x_0)=1}^{\infty}y^{n^2(x_0)}\cos\l[2\pi
    n(x_0)\phi(x_0)\r] \r\}\no\\
&=& \sum_{m\in N}\, y_m\,\cos\,[\,2\pi m\,\phi(x_0)\,],   \lb{vortexsum}
\ea
with
\ba
y_1&=&2y\,[\,1+y^2 + \Oh(y^2)\,],  \no\\
y_2&=&-y^2\,[\,1+2y^2+\Oh(y^3)\,], \no\\
y_3&=&(2/3)y^3+\Oh(y^5),     \no\\
y_4&=&y^4+\Oh(y^5),     \lb{y}
\ea
and $y_m (m\geq5)\sim \Oh(y^5)$.
In (\ref{y}) only a vortex with a single charge $(n(x_0)=\pm1)$
contributes to the most dominant terms in low-order harmonics,
$y_1,y_2$ and $y_3$.
Only in $y_4$ and higher-order harmonics, a doubly ($n(x_0)=\pm2$)
charged vortex provides the leading contributions.
Rescaling $\phi\ra\sqrt{\ka}\phi$ and identifying the sum $\sum_{x_0}$
with the continuum expression we get
\ba
Z_{\msG}&=&\prod_x \int\! \D\phi(x)
 \exp\l\{-\intx \l[\,{1\ov2}(\p\phi)^2+{M^2\ov2}\,\phi^2\r.\r.\no\\
 &&\l.\l.-\L^2\sum_{m\in N}\, y_m\,\cos(\,2m\pi\sqrt{\ka}\phi\,)\,\r]\,
\r\},   \lb{msG}
\ea
with $y_m$ given by (\ref{y}).
As in the lattice approach~\cite{JKS}, we have now reached the
massive s-G system starting from the frozen U(1) Higgs model.
Regarding each $y_m(\el)$ as independent coupling constants and
investigating their RG behaviors, we can perform the systematic
study of the dynamical problem: whether vortices are relevant or not
at long distances.

Before proceeding to the actual RG analyses let us give some remarks.

In ref.~\cite{JKS} Jones et al. have commented on the RG behavior of the
massive s-G theory derived from the lattice frozen U(1) Higgs model.
They argued in favor of the KT phase transition and predicted an
existence of the phase ($\ka<\ka_{\mbox{\scriptsize c}}$)
where the Higgs mechanism operates.
However, their argument is based on a hypothesis that in each of the
steps of RG the super-renormalizable mass term is irrelevant and can
be ignored.
This hypothesis may be plausible for the short-distance behavior
but not so for the long-distance behavior which determines
the phase structure of the model.
Classically a massive term is infrared relevant.
However, as is seen in the massless s-G theory
(pure XY model)~\cite{Kosterlitz},
a cosine operator which has the classical mass dimension zero,
could acquire large (=2) anomalous dimensions from quantum corrections
so that there can appear the (low-temperature) KT phase where vortices
are irrelevant at long distances ($y_m(\el\ra\infty)\ra0$).
Therefore, it is a non-trivial dynamical problem how the massive s-G
system (\ref{msG}) behaves at long distances.
To draw definite conclusions we must carefully investigate
the mixed renormalization-effects
among parameters $\ka, M^2$ and $y_m$.

For this purpose it is important that RG equations should be
constructed in terms of a recursion-relation method,
as has been done for the massless s-G theory in a real
space-time~\cite{Kosterlitz} and in a momentum
space~\cite{Wiegmann,Kogut}.
Since from the beginning we are working in the continuum with an
ultraviolet cut-off, it is natural to adopt the latter, i.e.~a
momentum-shell method.
It should be stressed that other conventional field-theoretical
schemes such as those based on {\it ultraviolet}
divergences~\cite{AGG} are {\it not}
suited for our purpose;
in such schemes the long-distance effect of super-renormalizable
terms such as a mass term could erroneously be missed.
This fact is actually exemplified in the standard O($N$) vector
non-linear $\si$ model with a {\it finite} external magnetic field
$h$, defined with the lagrangian
\b
\La={1\ov 2t}\l[ (\p \vec{\pi})^2+{(\vec{\pi}\cd\p\vec{\pi})^2\ov1-
\vec{\pi}^2} \r] + {h\ov t}(1-\vec{\pi}^2)^{1/2},
\lb{nls1}
\e
where $t$ is a coupling constant (temperature).
In the conventional scheme of dimensional renormalizations~\cite{BZ}
the last term is regarded only as an infrared cut-off and does not
affect the ultraviolet divergences (pole singularities at $d=2$).
As the result there appears no dependence on $h$ in the $\be$ function
for $t$.
This result is correct for the case with no external fields and also
for the case of non-zero $h$ in the critical region near the
ultraviolet fixed point.
The correct $\beta$ function for the non-zero finite $h$ was
first derived by Nelson and Pelcovits by use of the momentum-shell
recursion-relation method~\cite{NP}.
They discovered a non-trivial RG behavior revealing a crossover
phenomenon, which was completely missed in the field-theoretical
approach relying on short-distance singularities~\cite{BZ}.
The difference between these schemes also occurs in
other examples such as an anisotropic $\si$ model with no external
fields~\cite{NP}.

We think that these failures are not peculiar to a dimensional method
but could commonly happen in those utilizing ultraviolet
singularities to construct RG equations.
Thus it seems essentially important for us to perform the RG analysis
of the system (\ref{msG}) with the {\it recursion-relation} method,
along the original idea of Wilson~\cite{WK}.
This approach of the momentum-shell recursion has been applied to a
massless s-G theory~\cite{Wiegmann,Kogut} and it is not difficult to
employ it in the massive system (\ref{msG}).
\subsection{Construction of recursion equations for the s-G system}
\indent

As has been argued above we should adopt the momentum-shell
recursion-relation method to study correctly the long-distance
behavior of the massive s-G system (\ref{msG}).
The applications of the method to the massless version have already
appeared in refs.~\cite{Wiegmann,Kogut},
and we mainly follow the notations of ref.~\cite{Kogut}.
The RG of the system should start from the small-$y$
regime where a vortex excitation is fairly suppressed by a large chemical
potential ($-\ln y\gg1$) and so the harmonic terms can
be treated by a perturbation theory.
Since, as is seen in (\ref{y}), the higher-order harmonics possess the
coefficients of higher powers of $y$, we here consider only cosine
operators with coefficients to $\Oh(y^3)$, i.e.\  $y_1\sim y_3$.
Other higher-order ($m\geq4$) harmonics may also be relevant at the
long-distance limit but they are, at least initially, negligible
and does not have much influence on the RG behaviors of lower-order
harmonics ($y_1\sim y_3$).
If all the fugacities $y_m(\el)$ converge at $\el\ra\infty$ to zeroes,
then the system in the long-distance limit is described by a free
scalar boson with the mass $M^2=\ka/\be$, in complete agreement with
the prediction by a perturbation theory
of the original U(1) Higgs model (\ref{higgs}).
Our main interest then lies in whether there could exist the parameter
region in which such a Higgs picture is valid.
This question can legitimately be examined by a small-fugacity
$(|y_m|\ll1)$ perturbation theory.

The action of the massive s-G system consists of a free part $S_0$ of
the massive scalar field and of its harmonic term $S_{I}$ originated
from the vortex configurations in the U(1) Higgs model, i.e.
$S_{\msG}=S_{0}+S_{I}$ where
\ba
S_{0}&=&{1\ov2}\intx \l[\,(\p\phi)^2+M^2\,\phi^2\,\r], \lb{swaction}\\
S_{I}&=&-\L^2\,\sum_{m=1}\,y_m\, \intx \cos m\,\varphi_x\lb{voraction}
\ea
with $\varphi_x\equiv2\pi\sqrt{\ka}\phi_x\equiv2\pi\sqrt{\ka}\phi(x)$.
In the momentum-shell renormalization, the field $\phi$ is decomposed
into a low $\phi^{\prime}$ ($|p|<\L^{\prime}=\L-d\L$) and a high $h$
($\L^{\prime}<|p|<\L$) momentum parts
\b  \phi= \phi^{\prime}+h,  \lb{decomposition}\e
in which is assumed a smooth momentum-slicing procedure~\cite{WK,Kogut}.
Although this procedure is not explicitly specified here, it is in fact
argued in ref.~\cite{WK} that such a slicing procedure exists in
principle.
As is shown in refs.~\cite{Wiegmann,Kogut} the present treatment
applied to the massless s-G system in fact gives the same recursion
equations as those obtained by other methods including those based on
local singularities, at least to leading order~\cite{Kosterlitz,AGG}.
For the present {\it massive} s-G system defined at large distances
we rather believe that the momentum-shell recursion relation method
shall be more appropriate than others, by the reason we have remarked
in subsect.3.1.

Owing to the above field decomposition the partition function of the
system (\ref{msG}) reads
\ba
Z_{\msG}&=&\int\!\D\phi^{\prime}\exp\l\{-S_{0}(\phi^{\prime})\r\}\,
Z^{\prime}={\Cte}\int\!\D\phi^{\prime}\exp\l\{-F(\phi^{\prime})\r\},
\no\\
Z^{\prime}&=&\int\!\D h\, \exp\l\{-S_h-S_{I}(\phi^{\prime}+h)\r\}.
\lb{decopf}
\ea
Here $S_h\equiv S_{0}(h)$ is a free massive scalar-field action for the
high-momentum part $h$ and $F(\phi^{\prime})$ represents a free energy of
the low-momentum part $\phi^{\prime}$. Our task is to calculate
$F(\phi^{\prime})$ by a cumulant expansion with respect to $S_h$.
Explicitly,
\ba
F(\phi^{\prime})
&=&S_{0}(\phi^{\prime})\,-\,\ln(\,Z^{\prime}Z_h^{-1}\,)\no\\
&=&S_{0}(\phi^{\prime})\,+\,<\,S_{I}\,>
-\,(1/2)\,(\,<\,S_{I}^2\,> - <\,S_{I}\,>^2\,) \no\\
& &+(1/6)\,(\,<\,S_{I}^3\,>-3<\,S_{I}^2\,><\,S_{I}\,>+2<\,S_{I}\,>^3\,)
\no\\
                & & + \ldots,    \lb{cumexp}
\ea
where $S_{I}$ stands for $S_{I}(\phi^{\prime}+h)$ and the average
$<\;\;>$ is defined by
\b
<\,S\,>=\int\!\D h\,S\,\exp(-S_h)\,Z_h^{-1},\quad
Z_h=\int\!\D h\,\exp(-S_h).   \lb{RGvev}
\e

In Appendix A we have computed cumulants to $\Oh(y^3)$ and have
considered only the renormalizations of non-derivative $\cos m\varphi$
terms besides the $(\p \phi)^2$ operator.
The influence from other derivative operators generated in the first
order, are expected to be very small.
Combining (\ref{firstcum}), (\ref{A}), (\ref{secondcumu}) and
(\ref{thirdcumu}) we have the total free-energy density for the
low-momentum part $\phi^{\prime}$. It reads
\ba
\F
&=&(1/2)\l[\,1+2c\pi^2\de^2\De y^2_1d\el\,\r](\p\phi^{\prime})^2
   +(M^2/2){\phi^{\prime}}^2  \no\\
&&-\L^2[\,y_1-y_1\de\De(\,1-8\pi\de\De y_2+4\pi^2\de^2\De^2y_1^2\,)
   \,d\el\,]\,\cos\vp^{\prime}   \no\\
&&-\L^2[\,y_2-\de\De(\,4y_2-\pi\de\De y_1^2\,)
   d\el\,]\,\cos2\vp^{\prime}, \no\\
&&-\L^2\l[\,y_3 - \de\De\,(\,9y_3\,-\,8\pi\,\de\,\De\,y_1\,y_2
     \,+\,(4/3)\pi^2\de^2\De^2y_1^3\,)\,d\el\,\r]
     \,\cos3\vp^{\prime}                                 \no\\
&& + \,\Oh(y^4,(d\el)^2),                    \lb{Fdensity}
\ea
where $d\el\equiv\L^{-1}d\L=a^{-1}da,\;\de\equiv X+2\equiv\pi\ka,\;
\De\equiv(1+Z)^{-1},\;\vp^{\prime}\equiv2\pi\sqrt{\ka}\phi^{\prime}$
and $c$ is a constant number dependent on a momentum-slicing procedure.
By suitable renormalizations of the field and the parameters,
the above free-energy density can be rewritten in the original form
\b
\F={1\ov2}\,(\p\tilde{\phi^{\prime}})^2
   \,+\,{\tilde{M}^2\ov2}\,\tilde{\phi^{\prime}}^2\,
   -\,{\L^{\prime}}^2\,\sum_{m=1}^{3}\,\tilde{y}_m\, \cos m
   \sqrt{\tilde{\ka}}\,\tilde{\phi^{\prime}},
\lb{reFdensity}
\e
where the quantities with a tilde denote renormalized ones
for $\phi^{\prime},M$ and $y_m$.
Comparing (\ref{reFdensity}) with (\ref{Fdensity}) we can derive
the following approximate recursion equations:
\ba
{dX\ov d\el}&=&-{1\ov8}\,\de^3\De Y_1^2, \lb{X}\\
{dY_1\ov d\el}&=&\l[\,2\,-\,\de\,\De\,\l(\,1-{2\ov c}\,\de\,\De\,Y_2+{1\ov4c}\,
                 \de^2\,\De^2\,Y_1^2\,\r)\,\r]Y_1,   \lb{Y1}\\
{dY_2\ov d\el}&=&2\,Y_2\,-\,\de\,\De\,\l(\,4\,Y_2\,-\,{1\ov4}\,\de\,\De\,
                   Y_1^2\,\r),  \lb{Y2} \\
{dY_3\ov d\el}&=&2\,Y_3\,-\,\de\,\De\,\l(\,9\,Y_3\,-\,2\,\de\,\De\,
                 Y_1\,Y_2\,+\,{1\ov12}\,\de^2\,\De^2\,Y_1^3\,\r), \lb{Y3} \\
{dZ\ov d\el}&=&\l(\,2\,-\,{1\ov8}\,\de^2\,\De\, Y_1^2\,\r)\,Z,  \lb{Z}
\ea
where we have rescaled $4\pi\, c^{m/2}\,y_m=Y_m$.

\subsection{Solutions of recursion equations}
$\quad$
Having constructed the RG equations we now solve them numerically
and investigate the behavior of RG flows.
For simplicity we set $c=1$ in the numerical calculations.
(The choice of $c$ does not alter the qualitative
properties of RG flows.)
In the followings we fix the initial conditions for $Y_m$ by
$Y_1(0)=0.1$, which means from (\ref{y}) to take $y \simeq
4.0\times10^{-3}$.
Then, respecting other relations in (\ref{y}) provides $y_2 \sim - y^2
\sim - 1.6\times10^{-5}$ and $y_3 \sim (2/3)y^3 \sim 4.2\times
10^{-8}$, and the initial conditions for $Y_2$ and $Y_3$ are fixed
as
\b
Y_2(0)\,(=4\pi\, y_2)\,=-2\times10^{-4}, \quad
Y_3(0)\,(=4\pi\, y_3)\,=5\times10^{-7}.     \lb{Yini}
\e

Let us first see the solutions for the massless ($Z=0$) s-G
system to compare with those for the massive ($Z>0$) system.
Physically this system is related to the pure (non-gauged) XY model
or a two-dimensional Coulomb-gas system, and
the lowest-order solution of RG with only a single harmonics
($Y_1$) is already known \cite{Kosterlitz,Kogut}.
Here we will show the RG flows including the effects
of both higher-order corrections and higher harmonics.
The RG flows projected on the ($X,Y_1$) plane are depicted in
fig.5.
As is seen in this figure, their qualitative behaviors are the same as
those obtained by the lowest-order ($\Oh(y_1^2)$)
analyses~\cite{Kosterlitz,Kogut}.
The flow diagram exhibits the well known two-phase structure
distinguished by the KT phase boundary.
As is read from (\ref{X})$\sim$ (\ref{Y3}), the flow equations are
approximated by
\b {dX\ov d\el}\,\simeq\, -Y_1^2,\quad\quad {dY_1\ov d\el}\,\simeq\,
    -XY_1,\quad\quad{dY_2\ov d\el}\,\simeq\,{dY_3\ov d\el}\,\simeq 0,
\lb{XYap}
\e
in the neighborhood of $X\simeq Y_m \simeq 0$ where the phase boundary
in the $(X,Y_1)$ plane is described by $X\simeq Y_1$.
Figs.6(a) and (b) show the behaviors of $Y_1(\el),Y_2(\el)$ and
$Y_3(\el)$ as functions of a RG step $\el$, starting from the
representative points in the small ($X(0)=0.075$) and the large
($X(0)=0.125$) $X$ regimes.
It is observed that the higher harmonics ($m=2,3$)
behave similarly as the lowest harmonic ($m=1$), except in the
first small steps ($\ell\leq1$).
In the small-$X$ regime (fig.6(a)), $Y_2(\el)$ immediately
($\ell\ll1$) changes its sign to be positive and then starts
decreasing at $\el\sim0.62$.
Also $Y_3(\el) $ initially ($\ell\ll1)$ oscillates and
starts to decrease at $\el\sim0.74$.
All $Y_{m(=1\sim3)}(\el)$ continue to decrease until some
intermediate steps $\el=13.25\sim13.75$ where they turn  to increasing.
Whereas in the large-$X$ regime (fig.6(b)) all flows finally converge
to the $X$ axis ($Y_m=0$) which is a RG fixed line, although
$Y_2(\el)$ and $Y_3(\el)$ behave again non-trivially in initial
($\ell\leq1$) steps.

Since a single-charged ($n(x_0)=\pm 1$) vortex still gives the
most dominant contribution to the second and the third harmonics,
the above result that all $Y_{m(=1\sim3)}(\el)$ behave similarly at long
distances provides a consistent description for a vortex dynamics.
That is, the physical interpretation of the two phases given
in refs.~\cite{Kosterlitz,Kogut} persists to hold in our
extended system.
The large-$X$ (low-temperature phase of the XY model) regime
constitutes the KT phase where vortices are paired with a large
chemical potential (small fugacities $|y_m(\ell)|\ll1$) and so are
irrelevant at long distances.
The long-distance physics is described by a free massless
scalar field $\phi$ and is characterized by a power-law decay of its
correlation function $<\ex^{i\phi(0)}\ex^{-i\phi(x)}>$.
The non-gauged chiral four-Fermi model with large $N(>8)$
(\ref{phaselag}) exists in this phase.
In the small-$X$ (high-temperature) regime single vortices are
activated with a small chemical potential (large
fugacities $|y_m(\el)|\sim \Oh(1)$).
The power-law correlation of the massless scalar field is destroyed by
a vortex condensation and the disordered (vortex) phase is realized.

How are the above features of the massless s-G system are modified
if the mass term is added ($Z>0$), namely if the XY model is
gauged (the local $\Oh(2)$ model)?
The recursion equations (\ref{X})$\sim$(\ref{Y3}) show that
the limit $Y_m=0_{(m=1\sim 3)}$ is the fixed line of RG, irrespective
of the value of $Z$.
Thus, the most important question is whether there still exists the
region in which the RG flows run into this fixed line
where single vortices are completely suppressed and the long-distance
theory is described by a free massive scalar field (Higgs phase).

Fig.7 exhibits the behaviors of $Y_1(\el)$ for the three
initial mass conditions: $Z(0)=0,\; 0.01$ and $0.1$ with $X(0)=1.0$ and
$Y_1(0)=0.1$ fixed which corresponds, in the case of the massless
($Z=0$) system, to the initial condition set in the KT phase ($X\gg
Y_1$).
It is shown for the massive system ($Z>0$) that the decrease of
$Y_1(\el)$ stops at a certain $\el(=\el_{\c})$ the value of which
depends on an initial condition and that $Y_1(\el)$ turns to an
exponential increasing at large $\el(>\el_{\c})$.
Fig.8(a) shows the behaviors of $Y_1(\el)$ and $Z(\el)$ as
functions of a step $\el$, where $Z(0)$ is chosen to be $0.01$.
$Z(\el)$ remains small within small
steps $(\el<\el_{\c}\approx 2)$, where the second term (anomalous
dimensions of $y_1\L^2$) in the square bracket on the right-hand
side (r.h.s.) of the recursion equation (\ref{Y1}) is larger than or
comparable with the first term ($=2$, canonical dimensions of
$y_1\,\L^2$) and hence $Y_1(\el)$ decreases obeying the {\it quantum}
scaling as that for $Z=0$.
However, after some steps ($\el\sim \el_{\c}$) where $Z(\el)$ reaches to
be $\Oh(1)$, $Z(\el)$ grows large and reduces the anomalous dimensions
of $y_1\L^2$ significantly, $\Delta(\el\gg\el_{\c})\approx 0$.
As a result $Y_1(\el)$ turns to an exponential increasing according
to a {\it classical} scaling law (see also fig.8(b) plotted by a
logarithmic scale).
Fig.8(b) also indicates that this crossover and the scaling behaviors
in the massive system are common in three harmonics
$Y_m(\el)_{(m=1\sim3)}$, although in this figure is comprised
large-fugacity ($Y_m(\el)\gg1$) behaviors extrapolated from the
perturbative results.
We have checked numerically that these qualitative properties,
more or less, hold for all initial conditions set in a large $X$
regime.
There are no flows converging in the long-distance limit
to a $Y_m=0_{(m=1\sim3)}$ fixed line.
In any case including a small $X$ regime flows eventually
($\el\ra\infty$) go to the region where all $Y_m$ are
large.

For the massive s-G system, we thus conclude that there is no phase
transition nor a Higgs phase defined by $Y_m(\el\ra\infty)\ra0$.
Instead we have observed crossover phenomena from the
classical to the quantum scaling regimes in some intermediate
scales ($\el_{\c}$) which themselves depend on initial conditions.
These crossover phenomena are driven by an increasing of the
effective mass parameter $Z(\el)$($\approx$ gauge coupling).

Fig.8(a) also tells us that the effective probability of a vortex
in the generalized system defined by
$yP(\el)\approx(1/2)y_1(\el)P(\el)$ with $P$ in
(\ref{probability}) increases monotonically
and that the quantity $y\,n^*(\el)\approx(1/2)y_1(\el)n^*(\el)$ which
roughly measures the effective vortex density in the generalized
system
remains almost constant (increases slightly),
starting from the very dilute region $3.92\times10^{-4}$.
Therefore it is not obvious whether the diluteness of the vortex
distribution can actually be realized or not in the long-distance
limit ($\el\ra\infty$) even if it is chosen so initially ($\el=0$) at
some finite scales.
%
\section{Long-distance properties of the massive \newline
frozen U(1) Higgs model}
\indent

In this section we turn our attention to the theory in the
gauge non-invariant (anomalous) schemes ($b\neq0$).
The effective lagrangian is given by (\ref{eff})
with the gauge-field mass term proportional to $b$,
the anomaly parameter of the four-Fermi model.
We proceed again along the dual formalism.
\subsection{Double dual transformation and a coupled s-G system}
\indent

As in the previous case, performing the dual transformation
(\ref{dual}) and integrating over the gauge field
we obtain the following partition function for the system
$\La_{\Hig}+\La_{\ano}$:
\ba
Z&=&\prod_x\!\int\!\D\pho(x)\!\prod_x\!
    \int\!\D\ch(x)    \no\\
 & & \exp\l\{-\intx \l[\,{1\ov2\kap}\,(\p\pho)^2+{1\ov2\be}\,\pho^2
     +{\kaz\ov2}(\p\ch)^2-2\pi{\ka\ov\kap} i p \,\pho \,\r]\,\r\}
\lb{dualpfano}
\ea
with
\b \kap\equiv\ka+{b\,N\ov4\pi},\quad\quad
\kaz\equiv\ka\l(1-{\ka\ov\kap}\r).\lb{kapz}
\e
This system may be regarded as the XY model (the third term in the
exponent of (\ref{dualpfano})) coupled to a massive scalar field
$\pho$ through the vortex source
$p(x)=(2\pi)^{-1}\ep_{\mn}\p_{\mu}\p_{\nu}\ch(x)$.

The XY model lagrangian can be transformed by the following trick: we
introduce a fictitious gauge field $A^{\prime}$ and takes its zero
coupling limit $(e^{\prime}\ra 0)$ after performing the dual
transformation (\ref{dual}). Explicitly,
\ba
& &\exp\l\{- {\kaz\ov2} \intx (\p\ch)^2 \,\r\} \no\\
&=& \prod_x \!\int\!\D A^{\prime}(x)
    \exp\l\{-\intx \l[\,{1\ov4}(F^{\prime})^2
    +{\kaz\ov2}\,(\p\ch-e^{\prime}A^{\prime})^2 \,\r]\r\}
    \bigg|_{e^{\prime}\ra 0}  \no\\
&=& \prod_x \!\int\!\D A^{\prime}(x) \D\phz (x)  \no\\
& &  \exp\l\{-\intx \l[\,{1\ov2\be^{\prime}}\,\phz^2
    -i\phz\,(\ep\cd\p A^{\prime})
    +{\kaz\ov2}\,(A^{\prime}-\p\ch)^2 \,\r]\r\}
    \bigg|_{\be^{\prime}\ra\infty} \no\\
&=&\prod_x\!\int\!\D\phz (x)
   \exp\l\{-\intx \l[\,{1\ov2\kaz}\,(\p\phz )^2
  -2\pi i p \,\phz \,\r]\,\r\},
\lb{XYdual}
\ea
where $\be^{\prime}\equiv (e^{\prime})^{-2}$ and we have neglected
overall constant factors.
In the second equality of (\ref{XYdual}), we have rescaled
$e^{\prime}A^{\prime}\ra A^{\prime}$ and introduced a new
scalar field $\phz(x)$ via the dual transformation.
After the double dual transformation (\ref{dualpfano}) and
(\ref{XYdual}), the $\chi$ dependence in the effective action emerges
only through the vortex source operator $p(x)$.
Hence the $\chi$ configurations which can affect the partition function,
are only vortex configurations $\ch(x)=n(x_0)\,\theta(x-x_0)$
with arbitrary integer charges $n(x_0)$ and centers $x_0$,
as has been specified in (\ref{higgspf}) by the delta functional
as the gauge-fixing procedure for the gauge-invariant system.

Substituting (\ref{XYdual}) into (\ref{dualpfano}) and integrating
over $\chi$ with the above consideration,
we get the following total partition function:
\ba
Z&=&\prod_x \int\! \D\phz(x)\D\pho(x) \sum_{n}
    \exp\l\{-\intx \l[\,{1\ov2\kaz}(\p\phz)^2
    +{1\ov2\kap} \l(\,(\p\pho)^2 \r.\r.\r.\no\\
 && \l.\l.\l. +M_1^2\,\pho^2 \,\r)\,\r]
     + 2\pi i \sum_{x_0} n(x_0)
\l[\,\phz+\l({\ka\ov\kap}\r)\pho\,\r](x_0)\,\r\}
\lb{anoswvor}
\ea
with $M_1^2\equiv \kap/\be$.
We thus have seen that our effective theory $\La_{\Hig}+\La_{\ano}$
is equivalent to a system of {\it vortices interacting with both
massless ($\phz$) and massive ($\pho$) scalar fields}.

As in the previous case, in order to systematically study whether
vortex excitations are actually relevant or not at long distances,
we add a chemical potential term $(\ln y)\sum_{x_0}n^2(x_0)$ to
reduce the vortex activation ($y\ll1$).
Then taking the sum over vortex charges and identifying the sum over
vortex centers $\sum_{x_0}$ with the continuum integral $a^{-2}\intx$,
we obtain the following {\it coupled system of massless and massive
s-G fields}$\,$:
\ba
Z&=&\prod_x \int\! \D\phz(x)\D\pho(x)
    \exp\l\{-\intx \l[\,{1\ov2}(\p\phz)^2
    +\, {1\ov2}\l(\,(\p\pho)^2+M_1^2\,\pho^2 \,\r) \r.\r.\no\\
&& \l.\l.-\L^2 \sum_{m\in N} y_m
    \cos\l(\,2m\pi\sum_{j=0}^1\sqrt{\kappa_j}\phi_j\,\r)\,\r]\r\},
    \lb{coupledsG}
\ea
where
\b \kao\equiv \ka^2/\kap = \ka - \ka_0,  \lb{kao}\e
and the fields have been rescaled as
$\phi_0\ra\sqrt{\ka_0}\phi_0$ and $\phi_1\ra\sqrt{\ka_+}\phi_1$.
Here $y_m=y_m(y)$ is of order $\Oh\,(y^m)$.
Eqs.(\ref{dualpfano}),(\ref{kapz}) and (\ref{coupledsG}) indicate that
for the regularization scheme $b<0$ ($\kaz<0$) the theory is
non-unitary and we consider only the case $\kaz>0$ which is
realized by the $b>0$ schemes.

\subsection{Recursion equations for the coupled s-G system}
\indent
Using again the recursion-relation method, we now study the
long-distance properties of the coupled s-G system (\ref{coupledsG}).
As in the previous case (sect.3), although the original system
(\ref{anoswvor}) corresponds to $y=1$ it is interesting to consider
the very small $y(\ll1)$ regime where the perturbative RG method
is applicable.
Such an analysis persists us to speculate on whether the original
system ($y=1$) could at all possess the phase where vortices are
irrelevant at long distances $(y(\el\ra\infty)\ra0$),
although we cannot have direct access to the large $y$ regime.
In contrast to the previous system, vortices
in the current system interact with each other not only through
a short-ranged interaction due to $\phi_1$ but also through a
{\it long-ranged} one due to the {\it massless} field $\phi_0$.
We thus suspect that the long-distance behavior of the coupled
system (\ref{coupledsG}) should be different from that of the pure
(one-component) massive system, even {\it qualitatively}.

The procedure of constructing the recursion equations for the coupled
system is the same as that for the pure system, except that we have
now two independent dynamical fields.
They are decomposed into low and high-momentum parts
\b
\phi_j=\phi_j^{\prime}+h_j,   \quad\quad (j=0,1)          \lb{phideco}
\e
the latter of which is integrated over with an infinitesimal
slicing to generate recursion relations.
In the current analysis, we consider all operators appearing in
the $\Oh(y^2,\p^2)$ approximation.
As has been observed in the previous case, it is quite unlikely that
higher-order corrections in $y$ may cause the qualitative change
in the behavior in the $y\ll1$ regime.

Now, in the initial point ($\el=0$) of the RG transformation, we have
only the standard s-G lagrangian (\ref{coupledsG}) with two cosine
operators ($m=1,2$).
Then, extending the procedures developed in Appendix A for the pure
massive s-G system, we calculate the renormalized free energy based on
the cumulant expansion (\ref{cumexp}) with respect to the free part
$S_0$ of the action in (\ref{coupledsG}), in which the average
$<\;\;>$ is to be replaced by
\b
<S>=\prod_{j=0}^1\int\!\D h_j\,S\,\ex^{-S_h}\,Z_h^{-1},\quad
Z_h=\prod_{j=0}^1\int\!\D h_j\,\ex^{-S_h}   \lb{cRGvev}
\e
with $S_h\equiv S_0(h_0,h_1)$ being a free kinetic action of the
high-momentum parts $h_j$.

The effect of the first-order cumulant is merely to renormalize
the fugacity parameters $y_m$ of the original harmonics,
\b
<\,S_{I}\,> = - \L^2\, \sum_{m=1}^2\,y_m\,
               \l(\,1\,-\,m^2\,\delta\, d\el\,\r)\,
        \intx \cos m\,\vp^{\prime}\,+\Oh(\,(d\el)^2\,)  \lb{cfirstcum}
\e
with $\delta\equiv\sum_{j=0}^1\Delta_j\delta_j,
\;\Delta_j\equiv (1+Z_j)^{-1},\;Z_j\equiv \delta_{1j}M_1^2\L^{-2},
\;\de_j\equiv X_j+2\equiv\pi\ka_j$ and
$\vp^{\prime}\equiv2\pi\sum_{j=0}^1\sqrt{\kappa_j}\phi_j
\equiv\sum_{j=0}^1\vp_j$.
However, the second-order cumulant generates new operators
as well as the wave-function renormalizations.
Up to irrelevant \newline higher-derivative operators, it reads
\ba
&&(-1/2)\,\l(\,<\,S_I^2\,>\,-\,<\,S_I\,>^2\,\r)        \no\\
&=& \intx\, \l\{\, c\,\pi^2\delta y^2_1\,d\el\,
   \sum_{j=0}^1\delta_j\,(\p\phi_j^{\prime})^2
   -\,\pi\, y_1^2\delta^2 \L^{2}\,d\el\, \cos2\vp^{\prime}\r.\no\\
&& -\,(c\,\pi^2\,/\,8)\,y_1^2\,\delta\,d\el\,
   \sum_{j=0}^1\p^2\vp_j^{\prime}
   \sin 2\vp^{\prime}\no\\
&& \l. +\,2\,c\,\pi^2\,\sqrt{\dez\,\deo}\,\delta y_1^2\,d\el\,
  \p\phz^{\prime}\p\pho^{\prime}
  +\,\Oh((d\el)^2,\p^4)\,\r\}.        \lb{csecocum}
\ea
Therefore, in order to complete the $\Oh(y^2,\p^2)$ RG, we
must start from the generalized action $S=S_0+S_{I}^{\prime}$ with
the bare interaction
\b
S_I^{\prime}=\intx \l[\,-\,\L^2\sum_{m=1}^2 y_m\cos m\vp
             -\,{1\ov2}\,\sum_{j=0}^1\,w_j\p^2\vp_j\sin 2\vp\,
             +\,v\,\p\phz\p\pho\,\r],
\lb{generalaction}
\e
where $v$ and $w_j$ are new coupling constants. They are initially
($\el=0$) zeroes but are generated at $\el>0$ by the
$\Oh(y^2)$ perturbation, as is manifested in (\ref{csecocum}).
The second term in (\ref{generalaction}) produces new contributions
to the first-order cumulant
\ba
<\,S_I^{\prime}\,>
&=&\intx\,\l\{\,-y_1\L^2\l(1-\delta d\el\r)\cos\vp^{\prime} \r.\no\\
&&-\L^2\l[ y_2\,-\,\l(4y_2\delta\,+\,\pi^{-1}W\cd\delta\r)
   \,d\el\,\r] \cos 2\vp^{\prime}     \no\\
& &-\,{1\ov2}\,\l(1\,-\,4\delta d\el\r)\sum_{j=0}^1\,w_j
    \p^2\vp_j^{\prime}\sin2\vp^{\prime}\,
               +\,v\p\phz^{\prime}\p\pho^{\prime}  \no\\
& &\l. +\,\Oh((d\el)^2)\,  \r\}
\lb{genecfirst}
\ea
with $W\cd\delta\equiv \sum_{j=0}^1 W_j\delta_j\Delta_j$
and $W_j\equiv 4\pi w_j$.
In the $\Oh(y^2)$ approximation, the second-order cumulant
for the new action (\ref{generalaction}) is precisely the one
in (\ref{csecocum}).

The sum of (\ref{csecocum}) and (\ref{genecfirst})
is cast into the following renormalized free-energy density in the
$\Oh(y^2,\p^2\,)$ approximation:
\ba  \F(\phz^{\prime},\pho^{\prime})
&=&(1/2)\,\sum_{j=0}^1\,
   \l(1+2c\pi^2\delta\delta_j y^2_1d\el\r)
   (\p\phi_j^{\prime})^2\,+\,(M_1^2/2){\pho^{\prime}}^2  \no\\
&& -\,y_1\L^2\l(1-\delta d\el\r)\cos\vp^{\prime}   \no\\
&& -\,\L^2\l[\,y_2\,-\,\l(4y_2\delta+\pi^{-1}W\cd\delta
   -\pi y_1^2\delta^2\r)d\el\,\r] \cos2\vp^{\prime}   \no\\
&&-\,{1\ov2}\,\sum_{j=0}^1\,\l[\,w_j\,-\,\l(4w_j-(c\pi/4)
   y_1^2\r)\delta d\el\,\r]\p^2\vp_j^{\prime}
   \sin2\vp^{\prime}                                  \no\\
&& +\,\l(v+2c\pi^2\delta\sqrt{\dez\deo}y_1^2d\el\,\r)
     \p\phz^{\prime}\p\pho^{\prime}\,+\,\Oh((d\el)^2).
\lb{anoFdensity}    \ea

Under the same approximation follow the recursion equations:
\ba
{dX_j\ov d\el}&=&-\,{1\ov8}\,\delta\,\delta_j^2\, Y_1^2, \quad\quad
{dY_1\ov d\el}=\l(2-\delta\r)\,Y_1,   \lb{aXY1}\\
{dY_2\ov d\el}&=&2\,Y_2\,-\,\l(4\delta Y_2+4cW\cd\delta
                 -{1\ov4}\delta^2Y_1^2\r),   \lb{aY2}\\
{dZ_j\ov d\el}&=&\l(2-{1\ov8}\delta\delta_j
                       Y_1^2\r)\,Z_j,  \lb{aZ}  \\
{dV\ov d\el}&=&-\,{1\ov16}\l(\sqrt{\delta_1}-\sqrt{\delta_2}\r)^2
               \delta Y_1^2,\quad\;
{dW_j\ov d\el}= -\,\delta\l(4W_j-{Y_1^2\ov 16}\r),
                 \lb{aVWj}
\ea
where the coupling constants
are rescaled by $4\pi c^{m/2}y_m\equiv Y_m$ and $V\equiv v$.
If necessary, we can add the $\Oh(y^3)$ corrections to the above,
although computations are complicated.
As the relatively feasible case, we calculated the $\Oh(y_1^3)$
contribution to the renormalization of the $\cos\vp$ operator.
The result is reduced to the following replacements in the r.h.s.
of the second equation in (\ref{aXY1}):
\b
(\,2-\delta\,)\,Y_1\longrightarrow
\l[\,2- \delta\,\l(1+{1\ov4\,c}\delta^2Y_1^2\r)\,\r]\,Y_1,
\lb{y1third}
\e
the effect of which is negligible in the small $y_1$ regime.
\subsection{Solutions of recursion equations for the coupled s-G system}
$\quad$
Before presenting the full numerical solutions of
(\ref{aXY1})$\sim$(\ref{aVWj}), let us observe the analytical behavior
of (\ref{aXY1}) at some limiting cases.
To the leading $1/N$ order, the relations (\ref{beka}), (\ref{kapz}) and
(\ref{kao}) lead to
\b  \kaz \approx {N\ov 4\pi}\,{b\ov 1+b}, \quad\quad
\kao \approx {N\ov 4\pi}\,{1\ov 1+b},        \lb{kaN}    \e
and for the special choice $b=1$
\b (X\equiv) X_0=X_1={N\ov8}-2,  \lb{X01}   \e
on which initial conditions for $X_0$ and $X_1$
are identical and the equality is not altered by the RG (\ref{aXY1}),
that is $X_0(\el)=X_1(\el)$ for any step $\el$.

First, in the strong gauge-coupling limit $Z_1\ra\infty$
recursion equations in (\ref{aXY1}) are equal to those for
the pure massless s-G system in the second-order
approximation ((\ref{X}) and (\ref{Y1}) with $Y_2=Z=0$ and $Y_1^3\ra0$).
They describe the KT transition with the phase boundary $X_0=Y_1$
(fig.5).

Next, if $Z_1=0$
\footnote{Do not confuse this case with the non-gauged ($e=0$) limit
of the original four-Fermi model, where there is no anomaly from the
beginning. The non-gauged model ($\approx$ XY model) is related to
the one-component massless s-G system (sect.3).}
the system is a two-component  massless s-G system in which
recursion equations in (\ref{aXY1}) are reduced to
\b
{dX_j\ov d\el}=-\,{1\ov8}\,X_j^2\,(X_0+X_1+4)Y^2,\quad\quad
{dY_1\ov d\el}=-\,(X_0+X_1+2)Y_1.
\e
The critical line exists at $Y_1=0,\, X_0+X_1+2=0$ which
corresponds to $N=24$ in the large-$N$ approximation (\ref{kaN})
with an arbitrary $b>0$.
Fig.9 exhibits the RG flows ($b=1$) projected on a $(X, Y_1)$ plane.
The phase boundary around the critical point $(X=-1,Y_1=0)$
is written by
\b x^2-{y^2\ov8}=0   \e
with $x,y$ denoting small deviations from the critical point, i.e.
$X=-1+x,Y_1=y$.

Roughly speaking, if the mass parameter $Z_1(\el)$ of $\phi_1$ would
scale up at long distances ($\el\ra\infty$) as for the pure
massive s-G system (sect.3), the RG flows, starting from a very small
but a non-zero $Z_1(0)$, should exhibit a crossover from the
two-component massless s-G ($Z_1=0$) to the pure
 massless s-G ($Z_1\ra\infty$) systems.
The critical behavior in the long-distance limit shall thus be
described by the latter.

Fig.10 reveals the RG flows numerically solved for the coupled s-G
system with $b=1$.
They are projected on the $(X,Y_1)$ plane and the initial value of
$Z_1$ is taken to be $Z_1(0)=0.1$.
We have checked that $Z_1(\el)$ in fact grows large at long distances.
Accordingly, the crossover behavior can be seen as expected above.
Since $Z_1(0)$ is small the flows are initially obeyed by the
two-component massless s-G system ($Z_1=0$).
Beyond some steps where $Z_1(\el)\sim \Oh(1)$, the quantum
fluctuation due to $\phi_1$ starts to be frozen and the flows are
getting submitted to the one-component massless s-G system
($Z_1\ra\infty$).
As the result, the KT transition emerges with the critical point
$X_0=Y_m=0,$ and the KT-like phase exists in the $X_0(\infty)>0$
regime.
In the zero-th order approximation this KT regime means, $X_0(0)>0$,
i.e.
\b
b\,\pi\,\ka\,N - 8\,\pi\,\ka - 2\,b\,N > 0,    \lb{KT-like region}
\e
which reduces in the large $N$ limit to
\b
N\,>\,N_{\mbox{\scriptsize c}}\equiv8\,{1+b\ov b}.          \lb{KT-like regime}
\e
Note that this critical point $X_0=0$ does not depend on the value of
$c$, i.e. on the choice of a slicing procedure, as is the case in the
usual massless s-G system.
As is seen in fig.10, since the phase boundary in the $(X_0,Y_1)$
diagram is more close to the $Y_1$ axis than that for a pure
massless s-G system, the higher-order corrections (due to non-zero $Y_1$)
to the above estimation of the critical point
is smaller than those for the latter.

In the KT phase, vortices are irrelevant at long distances and
the spectrum in the long-distance theory consists of a free massless
 scalar field ($\phi_0$) and a free massive scalar field ($\phi_1$)
with the mass square
\b
{M_1^2(\el\ra\infty)\ov1-V^2(\el\ra\infty)},   \lb{freemass}
\e
which scales up at long distances.
In the $b\ra0$ limit the KT-phase disappears
($N_{\mbox{\scriptsize c}}\ra\infty$).
This consequence is consistent with that obtained for the pure massive
s-G system (sect.3) deduced from the gauge-invariant ($b=0$) scheme.

The similar critical behaviors are also seen for other coupling
constants $y_2$ and $w_j$.
Figs.11(a) and (b) show the RG behaviors of coupling constants
$Y_1$, $10^{2(3)}Y_2$ and $10^2W$ as functions of $\el$, in the vortex and
the KT phases respectively.
In theses figures we have chosen $b=1$
so that $V(\el)=0$ as seen from (\ref{aVWj}) and (\ref{kaN}).
In the vortex phase (fig.11(a)) all the coupling constants
are relevant at long distances, whereas they are irrelevant
in the KT phase (fig.11(b)).
\footnote{Although we plotted in fig.11(a) only the data
to $\el=12$, we checked that the flows continue to converge
to zero values at sufficiently long distances ($\el\sim 50$).}
%
%
\section{Summary and Concluding remarks}
$\quad$
Previous studies of the so called dynamical symmetry breaking
in the (gauged) four-Fermi models has mostly relied upon the
mean-field type treatments.
As a simple and typical case, the naive application of
$1/N$ expansion to the axially U(1)-gauged four-Fermi theory in $1+1$
dimensions predicts the dynamical Higgs phenomenon~\cite{GN} which,
in contrast to the non-gauged model~\cite{Root}, is stable at least
qualitatively against higher-order corrections.
In this article we have then addressed the problem whether such a
picture is really valid or not, and have investigated the possibility
that topological excitations (vortices) of the model could
be dynamically relevant to the long-distance properties.
The latter issue is important for the former general question because
topological excitations are non-mean-field-like objects and, if
relevant, usually work toward disordering the system and restoring
the symmetry broken by the mean-field ansatz, as is known in the
dilute-instanton-gas arguments.
This problem is of general importance, with no special regard to the
space-time dimensionality.

To simplify the problem we have first derived the large-$N$
long-distance effective lagrangian along with the Witten's
prescription in a non-gauged model~\cite{Witten}.
We have kept a global chiral-U(1) symmetry without assuming any v.e.v.
for the angle part $\chi$ of the axial-U(1) order-parameter
$\si+i\pi$, and have put a v.e.v. for the radial part the value
of which is determined by the large-$N$ gap equation.
The derived two-point effective action consists of the non-local part
which is (axially) U(1) gauge-invariant, and of the contact
mass term for the gauge field the coefficient ($b$) of which is arbitrary
and depends on a fermion-loop regularization scheme.

If the (axially) gauge-invariant scheme is chosen, one only obtains the
non-local part.
This part displays the dynamical Higgs mechanism in a
gauge-invariant fashion {\it provided} $\chi$ is restricted to be
single-valued.
The lowest-derivative part of this non-local action is a frozen
U(1) Higgs model with two parameters.
For this effective theory we have studied the long-distance
relevance of multi-valued (vortex) configurations of $\chi$.
With the help of the dual transformation, the partition function of
the frozen U(1) Higgs model with an ensemble of single-vortex
configurations has been shown to be equivalent with the system
of many vortices interacting with a massive scalar field.
This system has been further generalized to incorporate an additional
chemical-potential term of vortices $(\ln y) n^2(x_0)$ with a new
parameter $y$ which controls the activity of vortices.
Reducing $y$ by hand, we have examined the dynamical response to the
compulsive reduction of the vortex activation.
The small $y$ regime of the system is described by the massive s-G
system with small fugacities ($y_m\sim\Oh(y^m)$), to which we have
applied a perturbative recursion-relation method in a momentum space.
It has turned out that the (Higgs) phase characterized by
$y_m(\el\ra\infty)\ra 0$ for all $m$, corresponding to the KT phase in
the massless s-G system, does {\it not} exist in the massive system and
hence {\it vortices are always relevant at long distances}.
At long distances beyond a crossover scale ($\el=\el_{\c}$), we
have observed that quantum fluctuations are frozen due to a rapid
increasing of the effective mass parameter ($Z(\el)\gg1$) and that
the effective fugacities $y_m(\el)$ obeys a classical scaling (fig.8(b)).

Therefore it is strongly suggested that the frozen U(1) Higgs model
or the effective theory of the axially U(1)-gauged four-Fermi theory
in the gauge-invariant scheme, corresponding to the $y=1$ line
in the generalized scalar-vortex system, does {\it not} reach in the
long distance limit the fixed line ($y=0$) where vortices are completely
irrelevant and the Higgs mechanism naively operates (fig.12(a)).
Hence, due to a general disordering (symmetry-restoring) nature of
non-paired vortices, we shall {\it not} be allowed to expect the naive
(dynamical) Higgs picture to hold macroscopically.

If the (axially) gauge non-invariant regularization scheme is chosen
for the fermion loops, the effective theory of the original four-Fermi
model is the frozen U(1) Higgs model with a non-zero gauge-field mass
term.
{}From academic interests we have also studied the long-distance relevance
of vortex excitations in this model.
Utilizing the double dual transformation we have shown that the model
is equivalent to the system of vortices interacting with both massive
and massless scalar fields.
This system has been mapped in the similar way as above to
the coupled system of massless and massive s-G fields.
The recursion-relation analysis has been performed with the result
that a KT-type transition occurs at some critical value of the
Higgs coupling which is given to the zero-th order approximation
by (\ref{KT-like region}) in the small $y$ regime (fig.12(b)).
This analysis and the approximate argument given in Appendix B
suggest that in the gauge non-invariant schemes ($b>0$) the present
four-Fermi theory undergoes a KT-type phase transition at
\b
N=N_{\c}\approx {8(1+b)\ov b},          \lb{critical N}
\e
in the number $N$ of fermion species (fig.12(b)).
The large-$N$ ($> N_{\c}$) regime of the system constitutes a KT-like
phase where topological excitations are irrelevant and the
long-distance properties are characterized by a free
massless scalar field.
This phase disappears ($N_{\c}\ra\infty$) in the gauge-invariant limit
($b\ra0$) at which $\ka_0$ is zero and the massless field is frozen.
In this limit the theory is thus consistently continued to the
gauge-invariant scheme ($b=0$), and the combined results in both
gauge-invariant and non-invariant schemes provides a
unified view for the phase structure of the four-Fermi
model in general schemes $b\geq0$.

There are some points left to be clarified such as to determine
the macroscopic order parameter characterizing the phase structure.
To make the phase structure more precise and concrete
it would be necessary to consider the v.e.v. of a Wilson loop
operator~\cite{JKS} without relying upon a dilute-instanton-gas
approximation.
One unanswered technical problem in our recursion-relation analysis
is that we have assumed a smooth momentum slicing procedure
but have not specified it explicitly.
Although the general possibility of taking such a slicing is argued in
literatures~\cite{WK} it would be desirable to present it in the
explicit form.

In refs.~\cite{BHNH}, Banks et al. and Halpern studied the
bosonization of the non-gauged SU($N$) Thirring model.
It may be an interesting further direction to apply the method
to the present gauged model and to compare the results with those
obtained here.

Anyhow the analysis developed in this article presents the prototype
that the long-distance picture of the dynamical symmetry breaking
based on a naive mean-field treatment should be modified even
{\it qualitatively} due to the {\it topological} effects.
Note that {\it their existence itself is not special
to $1+1$ dimensions nor to the abelian nature of the gauge group}.
Then, it may be suspected that similar modifications of the
long-distance picture {\it could} occur as well in higher-dimensional
four-Fermi and Higgs models icluding those with non-abelian gauge
symmetries.
A semi-classical argument supporting this conjecture already
exists in a certain model~\cite{Polyakov,BMK}.
In the RG point of view one important difference between four and
lower than four dimensions may be that the gauge coupling constant
is dimensionless in the former while has positive mass dimensions
in the latter.
As is suggested from our analysis in the gauge non-invariant
scheme, the presence of a massless mode that causes the long-ranged
logarithmic interaction between instantons,
would be crucial for the long-distance properties of the system
in any dimensions.
We hope that the logical procedures developed here will supply
hints for the proper RG studies of the long-distance
{\it quantum} dynamics in more realistic models of particle physics.

Our results obtained for the gauge-invariant scheme may have
implications to some low-dimensional systems in solid state physics.
In the context of superconductivity mentioned in the
introduction, we suspect that in $2+1$ dimensions the Meissner
effect would not occur at finite temperature in cases where the
gauge field is to be treated {\it quantum} mechanically, which may
depend on experimental situations.
Although the KT-type phase transition observed in the gauge
non-invariant scheme is very interesting, we are not certain whether
such an {\it explicit} breaking mass term of the gauge field could be
relevant or not to real condensed-matter systems.

With the different motivation from ours, Ichinose and
Mukaida~\cite{IM} recently performed a recursion-relation analysis
of the massive s-G model with a single harmonic $(m=1)$ and drew the
similar conclusion to ours for the (lattice) frozen U(1) Higgs model.

The author would like to thank I. Ichinose, S. Iida, M. Kato, M.
Ninomiya and K. Shizuya for useful discussions and comments.
He is also grateful to S. Iida and members of YITP (Uji)
for their helps in his computer jobs.
\section*{Appendix A. $\;$ Evaluation of cumulants in the
recursion-relation analysis}
\setcounter{equation}{0}
\renewcommand{\theequation}{\mbox{\rm A.\arabic{equation}}}
\indent

In this appendix we present the calculation of cumulants appearing in
the recursion-relation analysis of the massive s-G system (\ref{msG})
in subsect.3.2.
Besides the original kinetic term $(\p\phi)^2$
we consider here only non-derivative harmonic operators $\cos m\vp$
within $\Oh(y^3)$ approximation.
There is no difficulty in extending the procedure to that applicable
for the coupled s-G system treated in subsect.4.2 where instead we
take into account all $\Oh(y^2,\p^2)$ terms consistently.

 The first-order cumulant $<S_{I}>$ is calculated easily with the
result
\b
<S_{I}>=-\L^2\sum_{m=1}y_m A^{m^2}(0)\intx\cos m\vp^{\prime}_x,
\lb{firstcum}
\e
where $A(x)=\ex^{-G_x}$ is defined with the Green function $G_x$ for $h$
\b
(2\pi^2\ka)^{-1}G_x=<h(x)h(0)>=\int_{\L^{\prime}<|p|<\L}\!d^2\!p\;
{\ex^{ip\cd x}\ov p^2+M^2}.   \lb{hgreen}
\e
To $\Oh(d\el=\L^{-1}d\L)$ we evaluate
\b
A^{m^2}(0)=1-m^2\pi\ka(1+Z)^{-1}d\el.  \lb{A}
\e

In the second-order cumulant we consider terms of
$\Oh(y_1^2)\sim\Oh(y^2)$ and of $\Oh(y_1 y_2)\sim\Oh(y^3)$
\ba
&&(-1/2)(<S_{I}^2>-<S_{I}>^2)\no\\
&=&(-1/4)y_1^2 A^2(0) \L^4 \intxy\,\no\\
&&  [\,(A_{xy}^2-1)\cos(\vp^{\prime}_x+\vp^{\prime}_y)+(A_{xy}^{-2}-1)
   \cos(\vp^{\prime}_x-\vp^{\prime}_y)\,]\no\\
&&+(1/2)y_1y_2 A^5(0) \L^4 \intxy  \no\\
&&  [\,(A_{xy}^4-1)\cos(\vp^{\prime}_x+2\vp^{\prime}_y)
  +(A_{xy}^{-4}-1)\cos(\vp^{\prime}_x-2\vp^{\prime}_y)\,]\no\\
&\simeq&(1/4)y_1 \pi^2\ka A^2(0)\L^2 \intxi \xi^2\,C_{\xi}\intz
        (\p \phi^{\prime}_z)^2 \no\\
&&+(1/2)y_1 y_2 A^5(0) \L^4 \intxi\intz
  [\,C_{\xi}(2+C_{\xi})\cos\vp^{\prime}_z
  +B_{\xi}(2+B_{\xi})\cos3\vp^{\prime}_z\,]  \no\\
&&-(1/4)y_1^2 A^2(0) \L^4 \intxi \,B_{\xi}\intz\cos2\vp^{\prime}_z,
  \lb{secocumu}
\ea
where $\xi\equiv x-y,\,z\equiv(x+y)/2,\,A_{xy}\equiv A(x-y),\,
B_{\xi}\equiv A^2(\xi)-1,\,C_{\xi}\equiv A^{-2}(\xi)-1$.
In the present approximation taken in subsect.3.2 we have neglected
in the last semi-equality higher-derivative terms such as
$(\p \vp^{\prime})^{2n}_{(n\geq2)},\;
(\p \vp^{\prime})^{2n}\cos m\vp^{\prime}_{(n\geq1)}$ and
$(\p \vp^{\prime})^{2n}\sin m\vp^{\prime}_{(n\geq1)}$.
To $\Oh(d\el)$ we evaluate
\ba
\intxi \,B_{\xi}&=&\intxi \,C_{\xi}={1\ov2}\intxi \,B_{\xi}^2
={1\ov2}\intxi \,C_{\xi}^2,   \no\\
&=&4\pi^3\ka^2(1+Z)^{-2}\L^{-3}d\L,   \lb{B}\\
\intxi \xi^2 \,C_{\xi}&=&4\,c\,\pi^2\ka(1+Z)^{-1}\L^{-5}d\L,  \lb{C}
\ea
where $c\equiv\int\!d\la \la^3 \tilde{J}_0(\la)$
is a dimensionless constant depending on a momentum-slicing
procedure~\cite{Kogut,WK}. Here $\tilde{J}_0(\la)$ denotes a Bessel
function modified so that $\la$ integral converges~\cite{Kogut}
according to a smooth momentum-slicing which we have not specified here.
Inserting (\ref{B}) and (\ref{C}) into (\ref{secocumu})
we obtain to $\Oh(d\el)$
\ba
&&(-1/2)(<S_{I}^2>-<S_{I}>^2)\no\\
&\longrightarrow&c\,\pi^4\ka^2y_1^2(1+Z)^{-1}d\el\intx\,
(\p\phi^{\prime})^2\no\\
&&+\,8\pi^3\ka^2y_1 y_2 (1+Z)^{-2}\L^2 d\el\intx\,
  (\cos\vp^{\prime}+\cos3\vp^{\prime})\no\\
&&-\,\pi^3\ka^2y_1^2(1+Z)^{-2}\L^2 d\el\intx \cos2\vp^{\prime}.
\lb{secondcumu}\ea

Similarly we evaluate the third-order cumulant within the same
approximation
\ba
&&(1/6)(<S_{I}^3>-3<S_{I}^2><S_{I}>+2<S_{I}>^3) \no\\
&=&{1\ov24}A^3(0)y_1^3 \L^6 \int\!\!\int\!\!\int\!d^2\!x\,d^2\!y\,d^2\!z\no\\
&&  [\,(A_{xy}^2A_{yz}^2A_{zx}^2-A_{xy}^2-A_{yz}^2-A_{zx}^2+2)
  \cos(\vp^{\prime}_x+\vp^{\prime}_y+\vp^{\prime}_z)\no\\
&&(A_{xy}^2A_{yz}^{-2}A_{zx}^{-2}-A_{xy}^2-A_{yz}^{-2}-A_{zx}^{-2}+2)
  \cos(\vp^{\prime}_x+\vp^{\prime}_y-\vp^{\prime}_z)   \no\\
&& +\,({\rm cyclic}\:{\rm permutations}\:{\rm in}\;x,y,z)\,]\no\\
&\simeq&{1\ov24}A^3(0)y_1^3 \L^6 \int\!\!\int\!\!\int\!
 d^2\!\xi\, d^2\!\eta\, d^2\!\!\rho  \no\\
&& [\,(B_{\xi}B_{\eta}+B_{\eta}B_{\xi+\eta}+B_{\xi+\eta}B_{\xi}+
     B_{\xi}B_{\eta}B_{\xi+\eta})\cos3\vp^{\prime}_{\rho} \no\\
&&  +\,3\,(B_{\xi}C_{\eta}+C_{\eta}C_{\xi+\eta}+C_{\xi+\eta}B_{\xi}
    +B_{\xi}C_{\eta}C_{\xi+\eta})\cos\vp^{\prime}_{\rho}\,]. \lb{thcum}
\ea
To $\Oh(d\el)$, a careful evaluation of $\xi,\eta$ integrals yields
\ba
&&(1/6)(<S_{I}^3>-3<S_{I}^2><S_{I}>+2<S_{I}>^3) \no\\
&\longrightarrow&-\,{1\ov3}A^3(0)y_1^3 \L^6\int\!\!\int\!d^2\!\xi\,
d^2\!\eta\, G_{\xi}G_{\eta} G_{\xi+\eta}\int\!d^2\!\!\rho\,
  (\cos3\vp^{\prime}_{\rho}+3\cos\vp^{\prime}_{\rho})\no\\
&\longrightarrow&-\,4\pi^5\ka^3y_1^3(1+Z)^{-3}\L^2 d\el\int\!d^2\!\!\rho\,
(\cos\vp^{\prime}_{\rho}+{1\ov3}\cos3\vp^{\prime}_{\rho}).  \lb{thirdcumu}
\ea
\section*{Appendix B. $\;$ An approximate argument for the $b>0$ system}
\setcounter{equation}{0}
\renewcommand{\theequation}{\mbox{\rm B.\arabic{equation}}}
\indent

We have seen in sect.4 that the coupled s-G system possesses
the KT phase at least in the region $X_0>0$ near $Y_m=0$.
Rigorously speaking, however, this only implies that
our effective lagrangian $\La_{\Hig}+\La_{\ano}$ {\it could}
be in the KT phase, since the original system (\ref{anoswvor})
corresponds to $y=1$ where the weak-coupling perturbation is not
valid.
Since the attractive fixed points on $Y_m=0$ for the coupled s-G
system exist for {\it any} non-zero value of $X_0$ and since we
have considered operators being most infrared-important in a RG sense,
it is plausible that our original model with a sufficiently
large $N$ is indeed in the KT phase.
Although it is difficult to give the proof, we will develop here
an approximate argument supporting partly this conjecture.
The idea is as follows.
Starting from the system (\ref{anoswvor}), we integrate over
all massive ($\pho$) scalar modes ($0<|p|<\L$).
In a certain region of parameters $e$ and $N$, the short-ranged
interaction among vortices due to $\pho$ can be neglected.
The system is then well approximated by that only of
a massless scalar field ($\phz$) coupled to
vortices with a sufficiently large chemical potential which is
generated by the $\pho$ fluctuation.
Without introducing any additional chemeical potential
the system is now described by the massless s-G system with a small
fugacity $y\ll1$, for which the perturbative RG analysis directly
applies.

First, the integration over all scalar fields $\phz$ and $\pho$ in
(\ref{anoswvor}) leads to the effective action
\b
2\pi^2
\sum_{x_0,y_0}\,n(x_0)
\l[\, \kaz D(x_0-y_0;0)+\kao D(x_0-y_0;M_1)  \,\r]\,n(y_0).
\lb{aSWvor}
\e
As in the pure XY model, an infrared divergence in $D(0;0)$ requires
a neutrality condition $\sum_{x_0}\,n(x_0)=0$.
Under this condition the effective action
 (\ref{aSWvor}) can be rewritten as
\ba
&&\l[\,2c_0\pi^2\kaz +{\pi\kao\ov2}\ln{\l(1+Z_1\ov Z_1\r)}\,\r]
   \sum_{x_0} n^2(x_0)   \no\\
&+&2\pi^2  \sum_{x_0\neq y_0} n(x_0)\l[\, \kaz D^{\prime}(x_0-y_0;0)
     +\kao D(x_0-y_0;M_1)  \,\r]\,n(y_0),           \lb{asveff}
\ea
where $c_0$ is a finite positive constant and $D^{\prime}$ is defined by
\ba
 D^{\prime}(x;0)&\equiv&D(x;0)- D(0;0)-c_0  \no\\
&\longrightarrow^{\raisebox{.8ex}
{\hskip -10mm\mbox{\scriptsize $|x|\L\ra\infty$}}}&
-{1\ov2\pi}\ln |x|\L. \lb{Dprime}
\ea
The terms in the first line of (\ref{asveff}) represents the chemical
potential of each vortex generated by $\phz$ and $\pho$ fluctuations,
and the second line expresses the vortex-vortex interaction due to
them.
As has been reviewed in subsect.3.1, the interaction ($D(x;M_1)$) due to
$\pho$  is short-ranged and decays exponentially
outside the range of $\Oh(Z_1^{-1})$.
On the other hand, from (\ref{asveff}) the probability to find a
vortex on a given position reads
\b
P=\exp (-2c_0\pi^2\kaz)\l({Z_1\ov 1+Z_1}\r)^{1+{X_1\ov2}}.   \lb{probs}
\e
Hence, the average number of other vortices lying in the
{\it $\pho$-interaction range} around a given vortex is evaluated to be
\b
n_1^* \sim \exp(-2c_0\pi^2\kaz)\, Z_1^{X_1/2}(1+Z_1)^{-(1+(X_1/2))}.
\lb{density1}
\e
For example, in the limit: (i) $Z_1\ra0$ with $X_1>0$ or
(ii) $X_1\ra\infty$ with $Z_1>0$,
both $P$ and $n_1^*$ vanish and the interaction (due to $\pho$)
can be neglected. \footnote{In the case (i), $Z_1$ must, however,
be kept positive.}
Therefore, in the large-$N$ and the very weak but non-zero
gauge-coupling region
\b
Z_1 \longrightarrow^{\raisebox{.8ex}
{\hskip -8mm\mbox{\scriptsize $N\ra\infty$}}}
{\hat{e}^2\ov 2\pi} \ll1
 \lb{weakcoup}
\e
with $\hat{e}\equiv e/\L$ being a dimensionless gauge coupling,
the system is approximated by that of a massless scalar field ($\phz$)
coupled to vortices with a large chemical potential
\ba
Z&=&\prod_x \int\! \D\phz(x) \sum_{n}
\exp\l\{-{1\ov2\kaz}\intx \,(\p\phz)^2
 -2\pi i \sum_{x_0} n(x_0)\phz(x_0) \r.\no\\
&&\l. + \ln y_0 \sum_{x_0} n^2(x_0)\,\r\},
\quad\quad y_0\equiv \l({Z_1\ov 1+Z_1}\r)^{\pi\kao\ov2}\ll1.
     \lb{effswvor}
\ea
This is equivalent to the massless s-G system with a small
fugacity ($Y_m\ll1$) where the perturbative RG analysis can
legitimately apply and predicts the KT transition to
occur at the critical point given by (\ref{KT-like region})
or (\ref{KT-like regime}).

\newpage
{\Large \bf Figure Captions}

\vspace{1cm}

{\bf Fig.1}: A $\si$ (a dashed line) tadpole diagram appearing in
              the second order gap equation of $1/N$ expansion, which
              includes an infrared divergence.
             The solid lines are fermion propagators and the wavy line
             represents the propagation of a massless NG ($\pi$) boson.

\vspace{0.3cm}

{\bf Fig.2}: $\rho$ (dashed lines) tadpole diagrams appearing in the
              second order gap equation of $1/N$ expansion, which
              contribute to infrared divergences.
             The solid lines are fermion propagators and the dotted
             lines represents the propagation of a massless KT
             ($\chi$) boson.
             Infrared divergences contained in both diagrams are
             cancelled with each other.

\vspace{0.3cm}

{\bf Fig.3}: $\rho$ (dashed lines) tadpole diagrams in the leading
              order of $1/N$ expansion.
             The solid line is a fermion propagator.

\vspace{0.3cm}

{\bf Fig.4}: Feynman diagrams which contribute to the ($A,\chi)$
              two-point effective action in the leading order of $1/N$
              expansion. The solid lines are fermion propagators.
              The wavy and dotted lines represent the gauge ($A$) and
              the KT ($\chi$) bosons.

\vspace{0.3cm}

{\bf Fig.5}: The RG flows for the massless ($Z=0$) s-G system,
              projected on a $(X,Y_1)$ plane.
              The dotted line is a phase boundary.

\vspace{0.3cm}

{\bf Fig.6(a)}: The RG behaviors of $Y_1(\el)$, $10^2Y_2(\el)$ and
                  $10^4Y_3(\el)$ as functions of a scale parameter $\el$,
                  in the small-$X$ $(X(0)=0.075)$ regime of the massless
                  s-G system.
\vspace{0.3cm}

{\bf Fig.6(b)}:  The RG behaviors of $Y_1(\el)$, $10^2Y_2(\el)$ and
                  $10^4Y_3(\el)$ as functions of a scale parameter $\el$,
                  in the large-$X$ $(X(0)=0.125)$ regime of the massless
                  s-G system.

\vspace{0.3cm}

{\bf Fig.7}: The RG behaviors of $Y_1(\el)$ as functions of a
              scale parameter $\el$, in the large-$X$ $(X(0)=1.0)$
              regime of the massless ($Z=0$: a solid line) and the massive
              ($Z(0)=0.01,\,0.1$: dashed and dotted lines) s-G systems.

\vspace{0.3cm}

{\bf Fig.8(a)}: The RG behaviors of $10Y_1(\el), Z(\el),
                 10^3y_1(\el)P(\el)$ and $10^3y_1(\el)n^*(\el)$ as
                 functions of a scale parameter $\el$, in the large-$X$
                 ($X(0)=1.0$) regime of the massive ($Z(0)=0.01$) s-G
                 system.

\vspace{0.3cm}

{\bf Fig.8(b)}: The RG scaling behaviors of $Y_1(\el), Y_2(\el), Y_3(\el)$
                 and $Z(\el)$ as functions of a scale parameter
                 $\el$, plotted by a logarithmic scale in the vertical
                 axis. Initial conditions are the same as those for
                 fig.8(a).

\vspace{0.3cm}

{\bf Fig.9}: The RG flows for the two-component massless s-G system
              ($b=1$), projected on a $(X,Y_1$) plane.
              The dotted line is a phase boundary.

\vspace{0.3cm}

{\bf Fig.10}: The RG flows for the coupled s-G system ($b=1$) with
                 $Z_1(0)=0.1$, projected on a  $(X,Y_1)$ plane.
                 The dotted line is a phase boundary.

\vspace{0.3cm}

{\bf Fig.11(a)}: The RG behaviors of $Y_1(\el), 10^2Y_2(\el)$
                 and $10^2W(\el)$ as functions of a scale parameter
                 $\el$, in the small-$X$  ($X(0)=-0.02$) regime of the
                 coupled s-G system ($b=1$) with $Z_1(0)=0.1$.

\vspace{0.3cm}

{\bf Fig.11(b)}: The RG behaviors of $Y_1(\el), 10^3Y_2(\el)$
                 and $10^2W(\el)$ as functions of a scale parameter
                 $\el$, in the large-$X$  ($X(0)=0.05$) regime of the
                 coupled s-G system ($b=1$) with $Z_1(0)=0.1$.

\vspace{0.3cm}

{\bf Fig.12(a)}: The schematic behavior of the expected RG flows
                 in the generalized scalar-vortex system with $b=0$.
                 The vertical line represents $y$ and the horizontal line
                 stands for other parameters $(\ka,\be,...)$ collectively.
\vspace{0.3cm}

{\bf Fig.12(a)}: The schematic behavior of the expected RG flows
                 in the generalized scalar-vortex system with $b>0$.
                 The vertical line represents $y$ and the horizontal line
                 stands for other parameters $(\ka_j,\be,...)$ collectively.

\end{document}